\documentclass[journal,comsoc]{IEEEtran}

\usepackage[T1]{fontenc}
\usepackage{amsmath}
\usepackage{ntheorem}
\usepackage{amssymb}
\newtheorem{theorem}{\noindent \textbf{Theorem}}

\usepackage{epsfig,latexsym,graphics}
\usepackage{graphicx}
\usepackage{subfigure}
\usepackage{cite}
\usepackage{url}
\usepackage{multirow}

\usepackage{amsmath}
\usepackage[cmintegrals]{newtxmath}

\begin{document}

\title{Weak Radio Frequency Signal Detection Based on Piezo-Opto-Electro-Mechanical System: Architecture Design and Sensitivity Prediction}

\author{Shanchi Wu, Chen Gong, Chengjie Zuo, Shangbin Li, Junyu Zhang, Zhongbin Dai, Kai Yang, Ming Zhao, Rui Ni, Zhengyuan Xu, and Jinkang Zhu

\thanks{This work was supported in part by National Key Research and Development Program of China under (Grant 2018YFB1801904), Key Program of National Natural Science Foundation of China (Grant No. 61631018), Key Research Program of Frontier Sciences of CAS (Grant No. QYZDY-SSW-JSC003), and Huawei Innovation Project. }
\thanks{Shanchi Wu, Chen Gong, Chengjie Zuo, Shangbin Li, Junyu Zhang, Zhongbin Dai, Kai Yang, Ming Zhao, Zhengyuan Xu, and Jinkang Zhu are with University of Science and Technology of China, Email address: \{wsc0807, jy970102, dzb123, wulikai\}@mail.ustc.edu.cn,\{cgong821, czuo, shbli, zhaoming, xuzy, jkzhu\}@ustc.edu.cn.}
\thanks{Rui Ni is with Huawei Technology, Email address: raney.nirui@huawei.com.}}

\maketitle

\begin{abstract}
We propose a novel radio-frequency (RF) receiving architecture based on micro-electro-mechanical system (MEMS) and optical coherent detection module. The architecture converts the received electrical signal into mechanical vibration through the piezoelectric effect and adopts an optical detection module to detect the mechanical vibration. We analyze the response function of piezoelectric film to an RF signal, the noise limited sensitivity of the optical detection module and the system transfer function in the frequency domain. Finally, simple on-off keying (OOK) modulation with carrier frequency $\textbf{1}$ GHz is used to numerically evaluate the detection sensitivity. The result shows that, considering the main noise sources in system, the signal detection sensitivity can reach around $\textbf{-118.9}$ dBm at bandwidth $\textbf{5}$ MHz. Such sensitivity significantly outperforms that of the currently deployed Long Term Evolution (LTE) system.
\end{abstract}

\begin{IEEEkeywords}
weak signal, radio frequency communication, sensitivity power level, piezo-opto-electro-mechanical system, piezoelectric fim, optical coherent detection.
\end{IEEEkeywords}

\IEEEpeerreviewmaketitle

\section{Introduction}

\IEEEPARstart{H}{igh} sensitivity signal detection of weak radio frequency (RF) signals is a long and ubiquitous challenge, crucial in radio astronomy, medical imaging, navigation, and communication \cite{bagci2014optical}. Due to rapid development of wireless communications, more compact amplifier, filter, oscillator and mixer circuits are being designed and delivered. Currently, due to wider usage of higher frequency for wireless communication, the faster signal attenuation also requires high-sensitivity signal detection to extend the communication range. Moreover, the communication system for future Internet of Things embraces lower transmission power to extend the battery life, which requires higher receiver-side detection sensitivity \cite{8345596, 8516297, 8550722, 8594699, 8625491, 8641430}.

On the other hand, the state-of-the-art signal detection operates using the components based on electro-mechanical interaction. The interaction between optics, electronics and mechanics can further increase the detection sensitivity. The main limit of signal detection sensitivity originates from all types of noise, while the conversion of mechanically mediated microwave and optical signals can theoretically be close to unit efficiency and zero noise temperature \cite{regal2011cavity, safavi2011proposal, taylor2011laser, barzanjeh2012reversible, wang2012using, tian2012adiabatic, tian2015optoelectromechanical}. Applying such systems to the nano-scale systems (limiting electromagnetic and displacement fields to submicron sizes) provides opportunities for enhancing the coupling strength and increase the receiver sensitivity \cite{midolo2018nano}.

From the general theory of electromagnetics, the control of the propagation and conversion of light can be achieved by changing the refractive index in a particular medium, or by changing the physical boundary between different media. The refractive index of materials can be changed in a variety of ways, such as electric field \cite{liu2015review} and temperature \cite{midolo2018nano}. The disadvantage of these methods lies in the weak variations. In contrast, the change of mechanical displacement has a significant effect on light control (e.g. the phase of the beam and the frequency of the light in cavity), and the movement of the machinery can be easily driven by electrostatic or piezoelectric force. The study of cavity optical mechanics \cite{kippenberg2008cavity, aspelmeyer2014cavity} shows that the nano-mechanical resonators can be strongly coupled with microwaves \cite{o2010quantum, teufel2011circuit, faust2012microwave} or optical fields\cite{groblacher2009observation, verhagen2012quantum}. Such characteristics increase the possibility for us to convert the RF signal to the optical signal. It is reported that high sensitivity nanometer opto-electro-mechanical system can be achieved by using high quality factor nano-film\cite{bagci2014optical}. Through the coupling inductor, the input microwave signal enters the LC circuit and is loaded on the film. Under the electrostatic force, the film vibrates and the laser phase incident on the surface changes. Consequently, the small film vibration can be detected at the back end. However, electrostatically driven nanoscale opto-electro-mechanical system has a low working frequency range (several Mega Hertz), which cannot meet the Giga Hertz frequency in RF communication systems. Therefore, a piezoelectric actuated opto-electro-mechanical system is desirable for RF frequency up to Giga Hertz. Although piezoelectric micro-electro-mechanical system (MEMS) resonators have been effectively used as frequency determining elements \cite{zuo20091, zuo2012cross, zuo2019hybrid}, there is no open report on using the vibration characteristics of piezoelectric MEMS for communication applications.

In this work, we propose an RF receiving system architecture based on the piezoelectric MEMS and optical coherent detection, called piezo-opto-electro-mechanical system (POEMS). To evaluate its feasibility in signal detection, we analyze the response function of piezoelectric film to an RF signal, and obtain the surface vibration equation. Based on the noise limited sensitivity of the optical detection module, the detection sensitivity to different frequencies signals is given. Furthermore, we characterize the transfer fucntion of POEMS in the frequency domain, based on the first order perturbation theory. Finally, considering noise signals from wireless channel, coupling circuit, piezoelectric film and optical module, we evaluate the performance of an on-off keying (OOK) modulation from bandwidths $1$ kHz to $10$ MHz with carrier frequency $1$ GHz. It is calculated that the detection sensitivity can be significantly improved, compared with that of the currently used communication system at bandwidths $3.75$ kHz and $5$ MHz. Such results envisage the promise of using POEMS for high-sensitivity signal detection.

The remainder of this paper is organized as follows. In Section II, we elaborate the proposed system model under consideration. In Section III, we investigate the response functions of each module and the entire system, and provide the noise characterzation and sensitivity limit. Numerical results of each module are given in Section IV. In Section V, we perform link simulation of our proposed system and evaluate the system performance. Finally, Section VI provides the concluding remarks.

\section{System Model}
\subsection{System Diagram}
\begin{figure}[!t]
	\centering
	\includegraphics[width = 3.5in]{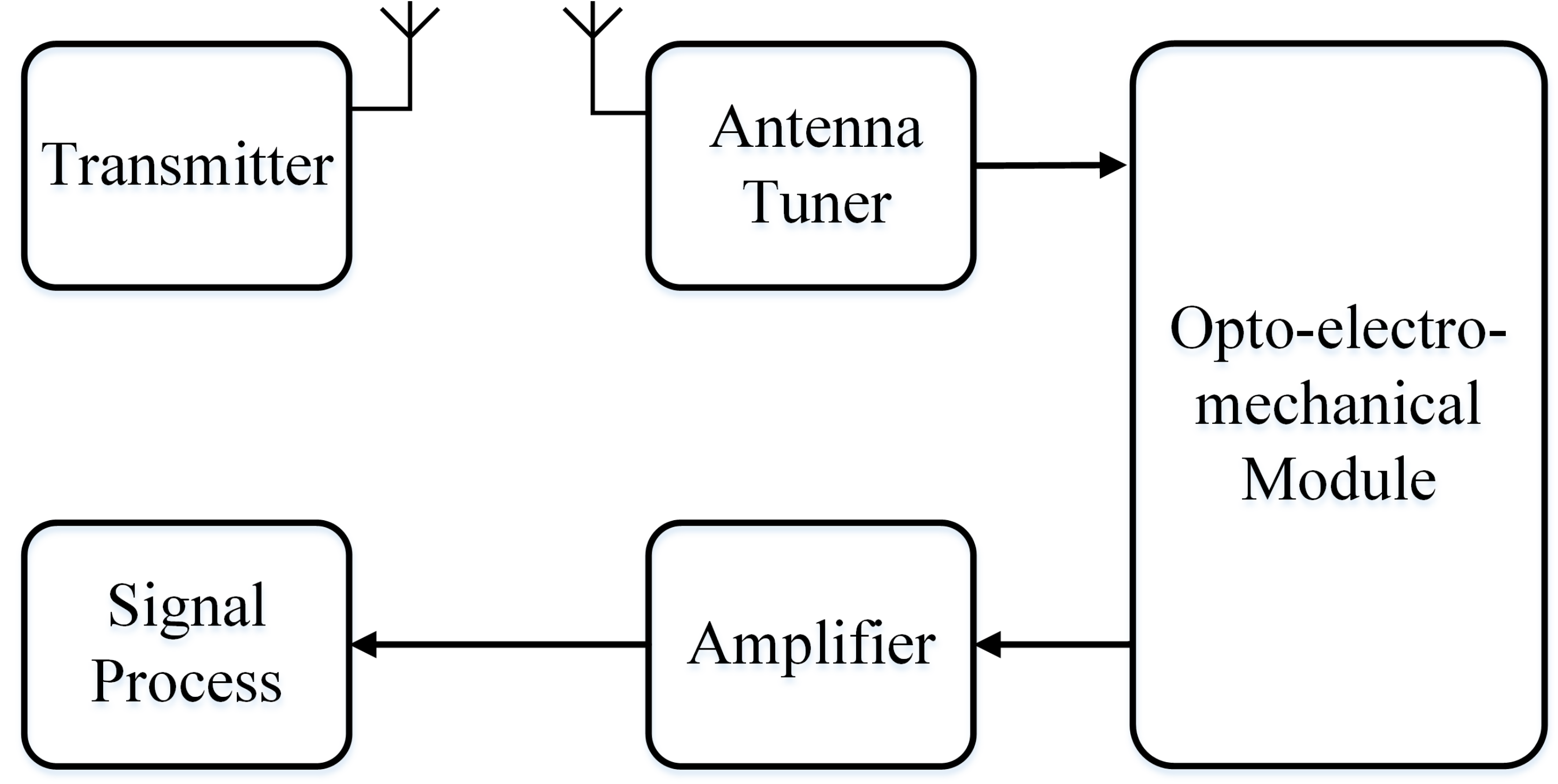}
	\caption{The diagram of the proposed system.}
	\label{fig.receiver}
\end{figure}

We consider a novel piezo-opto-electro-mechanical structure for weak power detection. The input signal is converted to mechanical vibration through the piezoelectric film, and the mechanical vibration is then converted to optical signal through an optical measurement system. The physical size of the piezoelectric film is in the order of submillimeter; and the band pass characteristics of piezoelectric film can realize frequency filtering effect. The diagram of the system under consideration is shown in Figure \ref{fig.receiver}.

\begin{figure*}[!t]
	\centering
	\includegraphics[width = 7in]{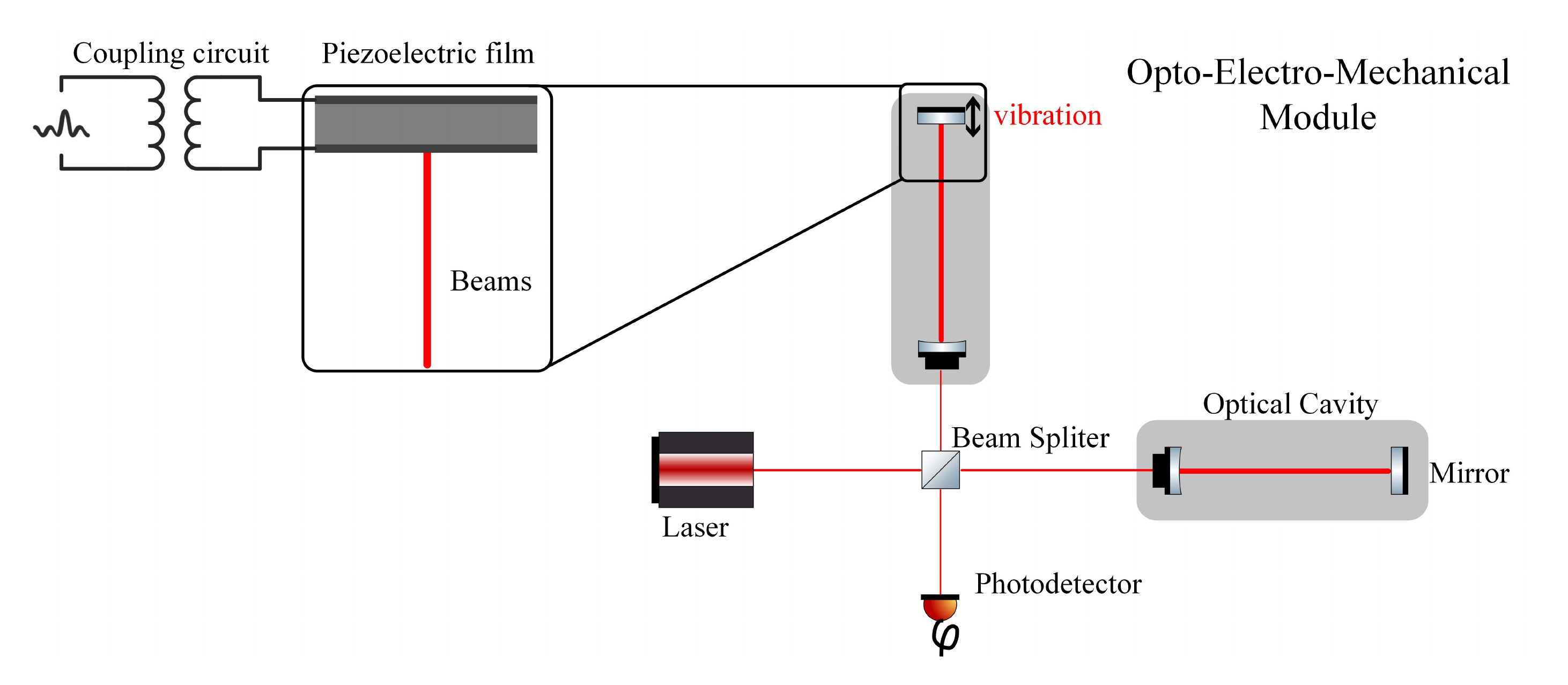}
	\caption{Diagram of the piezo-opto-electro-mechanical system.}
	\label{fig.oem_model}
\end{figure*}

Piezoelectric film coated with reflective surface acts as an "end mirror" of the optical detection module. When the piezoelectric film deforms, the end mirror position changes. The movement of piezoelectric film, which is driven by the RF signal, can be detected by the optical module output. As shown in Figure \ref{fig.oem_model}, the RF signal is coupled into the resonant circuit consisting of inductors and piezoelectric film. The film vibrates under the RF signal excitation, and the detected signal of the photodetector varies accordingly. The piezoelectric film and optical module will be characterized in detail in the subsequent subsections.

Such design can achieve high sensitivity by combining piezoelectric MEMS with the optical detection module. Due to the fundamental dynamic theory, electrostatic driven film oscillators cannot work at high frequencies, e.g., Giga Hertz. Piezoelectric actuated thin film oscillators can be adopted to solve such issue, such that high frequency signals can be detected with high sensitivity.

\subsection{Piezoelectric Film}
Piezoelectric materials have the property that an electric field applied in the direction of polarization can lead to the deform, due to the polarization effect of the dielectric. Aluminium nitride (AlN) is one of them, which has high thermal conductivity at low temperatures, good mechanical strength, high resistivity and corrosion resistance, and high resonant frequency. Thus, AlN resonators are attractive building blocks for electromechanical devices in micrometer and nanometer scales \cite{tonisch2006piezoelectric, zuo2010very, zuo2010multifrequency}.

The piezoelectric body can generate certain vibration mode activated by the external electric field. If an electric field is applied along a certian direction of the piezoelectric body, the non-zero piezoelectric constant associated with that direction can be employed to determine which vibration mode is likely to be excited. For weak RF signal reception, the thin film system has a linear response to external field strength. Although its three-dimensional system displacement is difficult to characterize, we adopt one directional responses for real applications, which approximately follows Hook law.

\begin{figure}[!t]
	\centering
	\includegraphics[width = 3.5in]{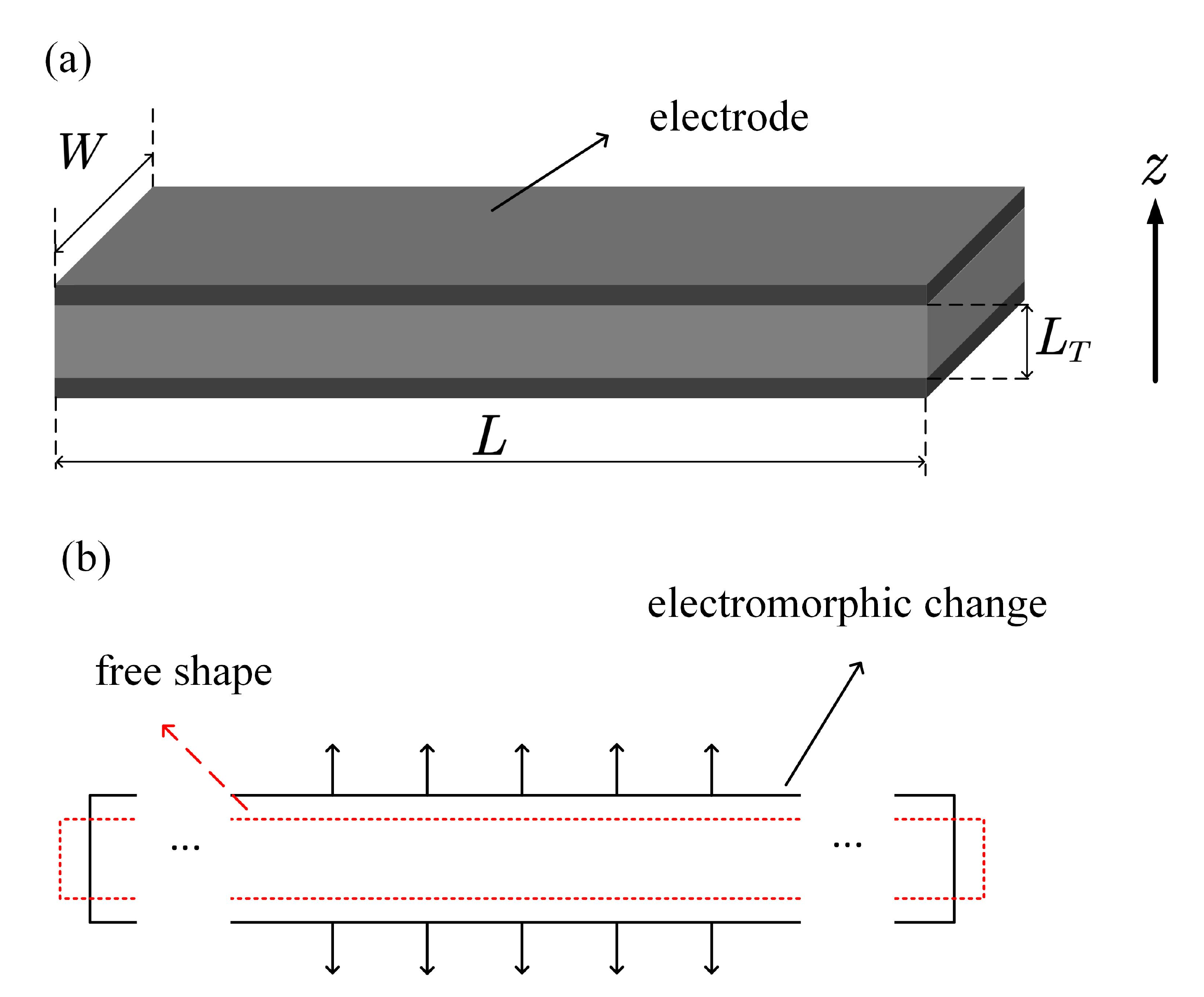}
	\caption{The sandwitch structure of piezoelectric film adopted, with length $L$, width $W$ and thickness $L_T$, along with the schematic diagram.}
	\label{fig.piezo_film}
\end{figure}

As shown in Figure \ref{fig.piezo_film} (a), the piezoelectric film adopted exhibits a sandwich structure, with an AlN thin film coated with metal layer as electrodes on the top and bottom surfaces, with length $L$, width $W$ and thickness $L_T$. Figure \ref{fig.piezo_film} (b) illustrates the shape in cross section, with and without electrical signal. In order to achieve vibration in the thickness direction and ignore the effects of vibration in other directions, the length and width of the film should be much larger than its thickness, i.e., $L, W \gg L_T$. It is important to note that the resonance frequency of the film (vibration in the direction of thickness) under consideration is determined by the thickness. Given piezoelectric material, the resonance frequency $f_0$ is inversely proportional to thickness. When a signal of certain amplitude with frequency $f$ is applied between two electrodes, the maximum deformation occurs at $f=f_0$.

\subsection{Optical Module}
Laser interferometer, as a type of precise optical measuring instrument based on the light interference, can measure the difference of optical paths generated by other certain relevant physical quantities. A typical laser interferometer consists of laser light source, splitting mirror, reflection mirror, polarizing optics and photoelectric detectors. Its basic structure is Michelson interferometer. Any change of optical path difference between two coherent beams will lead to the change of interference field (such as the movement of fringes, etc.), and the optical path change of a coherent beam is caused by the change of geometric path or refractive index of the medium through which it passes. The modern laser interferometer is based on the frequency stabilized laser with high stability of wavelength, whose measurement accuracy is significantly higher than that of other measurement methods. Optical interferometer has been adopted for measurement in astronomy, optics, engineering surveying, oceanography, seismology, spectrum analysis, quantum physics experiment, remote sensing and radar. The prototype of Laser Interferometer Gravitational-Wave Observatory (LIGO) system is a Michelson optical interferometer \cite{AbramoviciLIGO}.

We design the optical module to realize an optical interferometer, which utilizes the superposition of optical wave to obtain the phase information. As the vibration amplitude of piezoelectric film is tiny under the weak RF signal stimulation, high-precision displacement measurement is required, based on the optical coherent detection following the idea of Michelson interferometer. When a small shift of the end mirror occurs, the signal strength of the photodetector changes accordingly, caused by the phase change of the light in optical arm. In order to achieve higher precision, larger laser power and a longer light arm is desirable. In order to decrease the receiver module size, optical cavity is adopted, which significantly increases the measurement precision \cite{freise2010interferometer}.

\section{Sensitivity Analysis} \label{Sensitivity_Analysis}
\begin{figure}[!t]
	\centering
	\includegraphics[width = 3.5in]{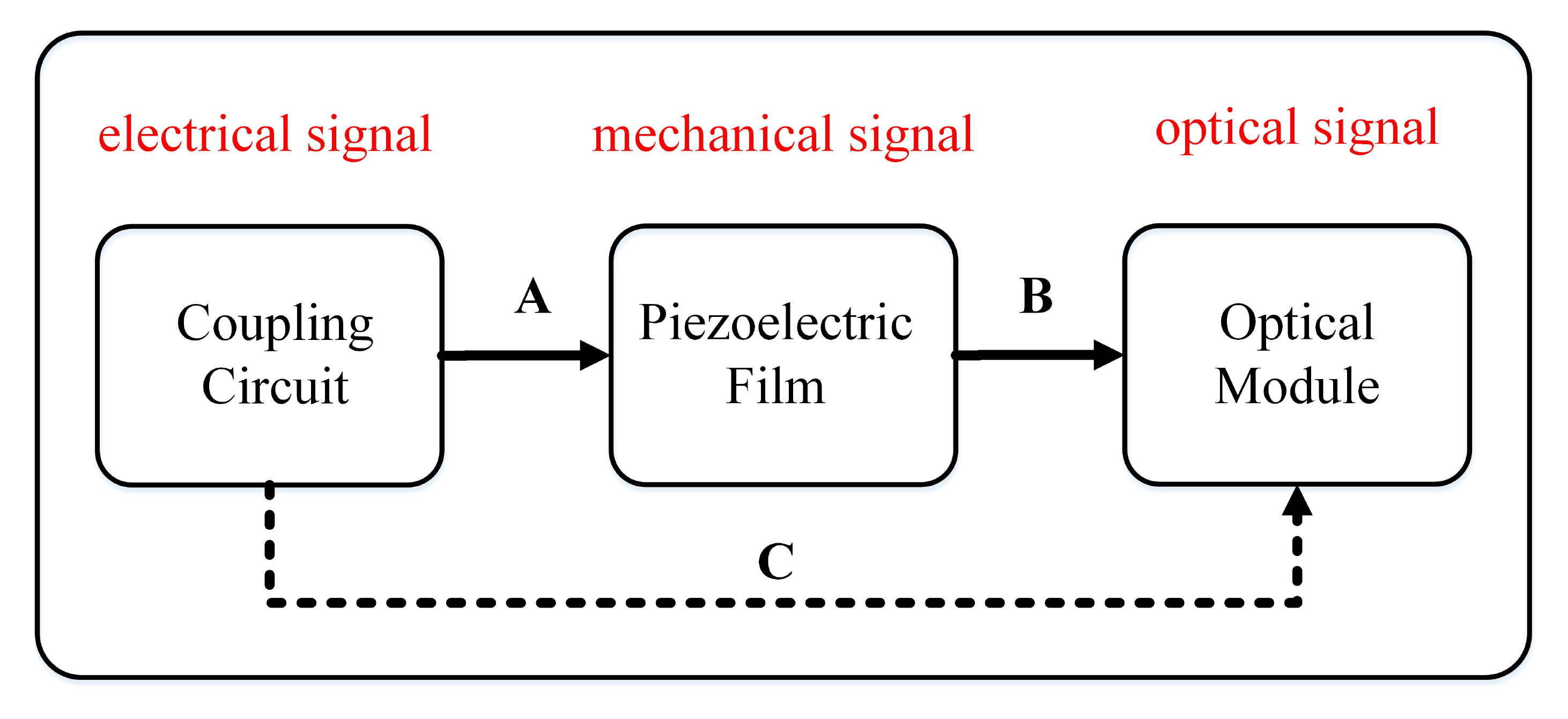}
	\caption{Diagram of the sensitivity analysis flow.}
	\label{fig.theoretical_block}
\end{figure}

We analyze the response of the piezoelectric film and the optical measurement module, based on which the system tranfer function in the frequency domain can be obtained. As a major system degradation factor, the noise components from film, circuits and optical module are characterized. Combined with the tranfer function, the sensitivity of the receiver modules can be obtained. The analysis will follow the blocks as shown in Figure \ref{fig.theoretical_block}. Firstly, we give the response of piezoelectric film to signals, which converts electrical signal  to mechanical displacement. Secondly, we investigate the response of optical module to mechanical vibration, which converts mechanical signal to optical phase. Then, based on the discrete modules, we can obtain the system transfer function from electrical signal to optical phase. Finally, considering the noise source, the noise limited sensitivity is analyzed.

\subsection{Response of Piezoelectric Film to Signals}\label{Response to excitation}
The alternating voltage is applied to the piezoelectric vibrator, and the mechanical vibration of the pieozoelectric vibrator will be excited by the inverse piezoelectric effect coupling, which will generate strain in the vibrator. Moreover, the mechanical vibration of the vibrator will generate current through the positive piezoelectric effect and feedback back to the circuit. When the applied driving voltage frequency is close to the eigen mechanical resonance frequency of the oscillator, the resonance will lead to large mechanical vibration amplitude. The overlap of the driving current and the feedback current may increase the current flowing through the oscillator. The impedance and admittance frequency characteristics of piezoelectric vibrators close to resonant frequencies can be approximated by an equivalent circuit.

Assume that the film length and width is significantly larger than the thickness, i.e., $L, W\gg L_T $. The thickness direction is the main factor. Thus, only the effect of stress component $T_z$ is considered, and other stress components can be ignored. As the electrode surface is perpendicular to the z axis, only the effect of electric field component $E_z$ is considered, and other electric field components can also be ignored. In addition, because only the edges of the film are fixed, the centeral part is free, i.e., the boundary condition of two surfaces in z-axis is free. The stress component $T_z$ is equal to zero, and the electrode surface is equipotential. In this case, we can choose $T_z$ and $E_z$ as independent variables and use the first class of piezoelectric equations as follows,
\begin{equation}
	\begin{aligned}
		T_z &= c^ES_z-e_{33}E_z,\\
		D_z &= e_{33} S_z+\varepsilon ^SE_z,
	\end{aligned}
\end{equation}
where $T_z, S_z, E_z$ and $D_z$ represent stress field, strain field, electric field and electric displacement vector components along the z-axis, respectively; $c^E$ is the elastic modulus at constant electric field strength; $\varepsilon^S$ is the relative permittivity at constant strain field strength; and $e_{33}$ is piezoelectric constant decided by material \cite{arnau2004piezoelectric}. Typically, $c^E$ and $\epsilon^S$ are treated as constants. Here, all vector variables are expressed as scalars since we only consider one-dimensional vibration. For convenience, in the following derivations, the letter subscript indicating the $z$ direction is neglected, shown as follows,
\begin{equation}\label{piezo_equation}
	\begin{aligned}
		T&=c^ES-e_{33}E,\\
		D&=e_{33} S+\varepsilon ^S E.\\
	\end{aligned}
\end{equation}
 
Letting $u$ represent the particle displacement in z-axis, given that the displacement of the piezoelectric film is treated as an elastic harmonic oscillator, we have
\begin{equation}
	\begin{aligned}
		\frac{\partial ^2u}{\partial t^2} &=\frac{c^D}{\rho}\frac{\partial ^2u}{\partial z^2},\\
		c^D &=c^E+\frac{e_{33}^2}{\varepsilon ^S},\\
	\end{aligned}
\end{equation}
where $\rho$ is the mass density of the film and $c^D$ is the elastic modulus at constant electric displacement vector \cite{rosenbaum1988bulk}.

Considering the separation of variables, i.e., $u\left( z,t \right) =Z\left( z \right) \cdot e^{j\omega t}$, and defining phase velocity $v_a \triangleq \sqrt{\frac{c^D}{\rho}}$ of the vibration module along the z-axis, the wave equation $Z(z)$ is given as follows,
\begin{equation}
	\begin{aligned}
		\omega ^2Z+\frac{c^D}{\rho}\frac{d^2Z}{dz^2} &=0,\\
		\frac{d^2Z}{dz^2}+\frac{\omega ^2}{v_{a}^{2}}Z &=0.\\
	\end{aligned}
\end{equation}

\begin{theorem}\label{theorem_solution_no_damping}
	Under excitation signal $V=V_0\cdot e^{j\omega_r t}$ and free boundary condition $T_{z=0}=T_{z=L_T}=0$, the surface vibration at $z=L_T$ is given by
	\begin{equation}
		u\left( L_T, t \right) =\frac{e_{33} D_0}{\varepsilon ^S c^D \beta}\tan \left( \frac{\beta L_T}{2} \right) \cdot e^{j\omega_r t},
	\end{equation}
	where $D_0$ is the amplitude of electric displacements vector; $\beta \triangleq \frac{\omega_r}{v_{a}}$ is a defined coefficient; and $\omega_{r}$ is the resonant frequency.
\end{theorem}

\begin{IEEEproof}
	Please refer to Appendix \ref{solution_no_damping}.
\end{IEEEproof}

In a real system, the relationship between force and strain may not be linear. Accordingly, parameters such as damping coefficient $\eta$ are introduced to characterize the new system, which can be written as $T =c^DS+\eta \frac{dS}{dt}$.

Under the condition that the displacement of the piezoelectric film is treated as a damped elastic harmonic oscillator, the wave equation satisfied by the displacement in damping system is given by
\begin{equation}
	c^D \frac{\partial ^2v}{\partial ^2z}+\eta \frac{\partial ^3v}{\partial t\partial ^2z} =\rho \frac{\partial ^2v}{\partial t^2},
\end{equation}
where $v(z,t)={\partial u(z,t)}/{\partial t}$ represents the particle velocity; $\rho$ represents the mass density of the film\cite{rosenbaum1988bulk}.

\begin{theorem}\label{theorem_solution_damped}
	For piezoeleteic film system with damping coefficient $\eta$, under the excitation signal $V=V_0\cdot e^{j\omega_s t}$ and boundary condition $T_{z=0}=T_{z=L_T}=0$ , the surface vibration at $z=L_T$ is given by
	\begin{equation}\label{Eq7}
		u\left( L_T,t \right) \approx -j\frac{4QV_0e_{33}}{\pi ^2 c^D}\cdot e^{j\omega_s t},
	\end{equation}
	where $e_{33}$ is the piezoelectric constant and $Q={c^D}/{\eta\omega_s}$ is the quality factor; and $\omega_{s}$ is the series resonance frequency.
\end{theorem}
\begin{IEEEproof}
	Please refer to Appendix \ref{solution_damped}.
\end{IEEEproof}

\subsection{Response of Optical Module to Piezoelectric Film Vibration}
\begin{figure}[!t]
	\centering
	\includegraphics[width = 3.5in]{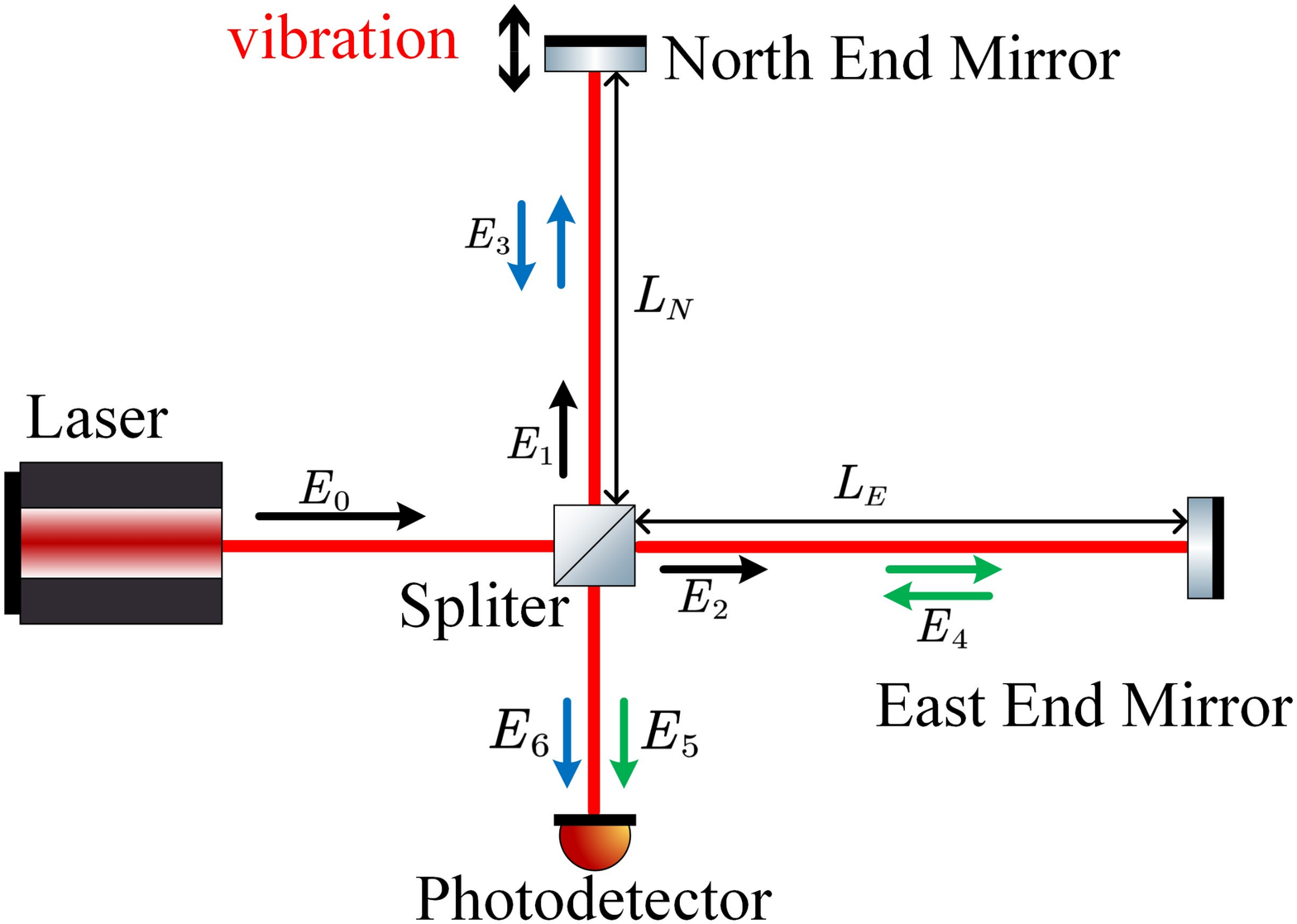}
	\caption{Diagram of Michelson interferometer.}
	\label{fig.optical_model}
\end{figure}

For classical Michelson interferometer, the beam propagation can be represented by electric field, as shown in Figure \ref{fig.optical_model} \cite{freise2010interferometer}. Assuming that all parameters of the optical components are known, the output of optical field is characterized as follows,
\begin{equation}
	\begin{aligned}
		\mathbf{E_{out}} &= \mathbf{E_5} + \mathbf{E_6} \\
		&=\mathbf{E_0}rt\left( e^{\text{i}\left( \varphi _t+\varphi _{r1}+\Phi _1 \right)}+e^{\text{i}\left( \varphi _t+\varphi _{r2}+\Phi _2 \right)} \right),\\
	\end{aligned}
\end{equation}
where $\varphi _{r1},\varphi _{r2}$ and $\varphi_t$ are phase differences caused by reflection of upper surface, lower surface and transmission of the mirror, respectively; $r$ and $t$ are the reflectivity and transimittivity of end mirrors, respectively; $\Phi_1$ and $\Phi_2$ are phase differences caused by optical path differences in arms, respectively; and  $R=r^2$ and $T=t^2$. 

In a real system, we can only perform direct detection on the optical power or intensity. It is difficult to obtain the optical signal frequency of the commonly used laser directly through the photoelectric sensor. A common solution is to detect the beat of the coherent signal, which is the envelope of the intensity of two overlapping and coherent fields. For the output light field with two frequency components $\boldsymbol{E}=E_0\left[ \cos \left( \omega _1t \right) +\cos \left( \omega _2t \right) \right] $, the corresponding output power is
\begin{equation}
P = E_{0}^{2}\left( \cos ^2\left( \omega _1t \right) +\cos ^2\left( \omega _2t \right) +\cos \left( \omega _+t \right) +\cos \left( \omega _-t \right) \right),
\end{equation}
where $\omega _+=\omega _1+\omega _2$ and $\omega _{\_}=\omega _1-\omega _2$. 

For any interferometer systems, the optical arm length variation will affect the laser propagation in space and cause the phase variation. In order to analyze the influence of the moving end mirror amplitude and frequency on the output sideband signal, we apply periodic modulation signal $x_m =  a_s cos(\omega_s t + \varphi_s)$ to the optical arm, as shown in Figure \ref{fig.optical_model}. We have the following results on the output signal of the light field. The output signal of the light field after the reflection signals from the N-end mirror and E-end mirror, combined by the beam splitting mirror, is given by
\begin{equation}
	\begin{aligned}
	\mathbf{E_{out}} 
	&=\mathbf{E_0}rt\left( e^{\text{i}\left( \varphi _t+\varphi _{r1}+\Phi _1 + 2 k_0 x_m\right)}+e^{\text{i}\left( \varphi _t+\varphi _{r2}+\Phi _2 \right)} \right)\\
	&=\mathbf{E_0} rt\left( e^{\text{i}\left( \varphi _t+\varphi _{r1}+\Phi _1\right)} \left(1 + k_0 a_s (s^+ + s^-)\right) \right.\\
	& \left. \quad +e^{\text{i}\left( \varphi _t+\varphi _{r2}+\Phi _2 \right)} \right),\\
	\end{aligned}
\end{equation}
where $E_0$, $k_0$ and $\Delta L$ are the amplitude, wave vector and optical arm length differences of the carrier field, respectively; $s^+=e^{i(\omega_s t + \varphi_s + \pi/2)}$ and $s^-=e^{-i(\omega_s t + \varphi_s - \pi/2)}$ are upper and lower signal sidebands, respectively \cite{freise2010interferometer}.

\begin{theorem}\label{theorem_output_current}
	Assuming that the injection light power is $P_0$ and photoelectric responsivity of the photodetector is $\alpha (A/W)$, the output current of the Michelson interferometer is given by
	\begin{equation}
	\begin{aligned}
		I_{out} &= {\alpha P_0} \left[ cos^2(k_0 \Delta L) + k_0 a_s sin(2 k_0 \Delta L)cos(\omega_s t + \varphi_s) \right].
	\end{aligned}
	\end{equation}
\end{theorem}
\begin{IEEEproof}
	Please refer to Appendix \ref{output_current}.
\end{IEEEproof}

\begin{figure}[!t]
	\centering
	\includegraphics[width = 3.5in]{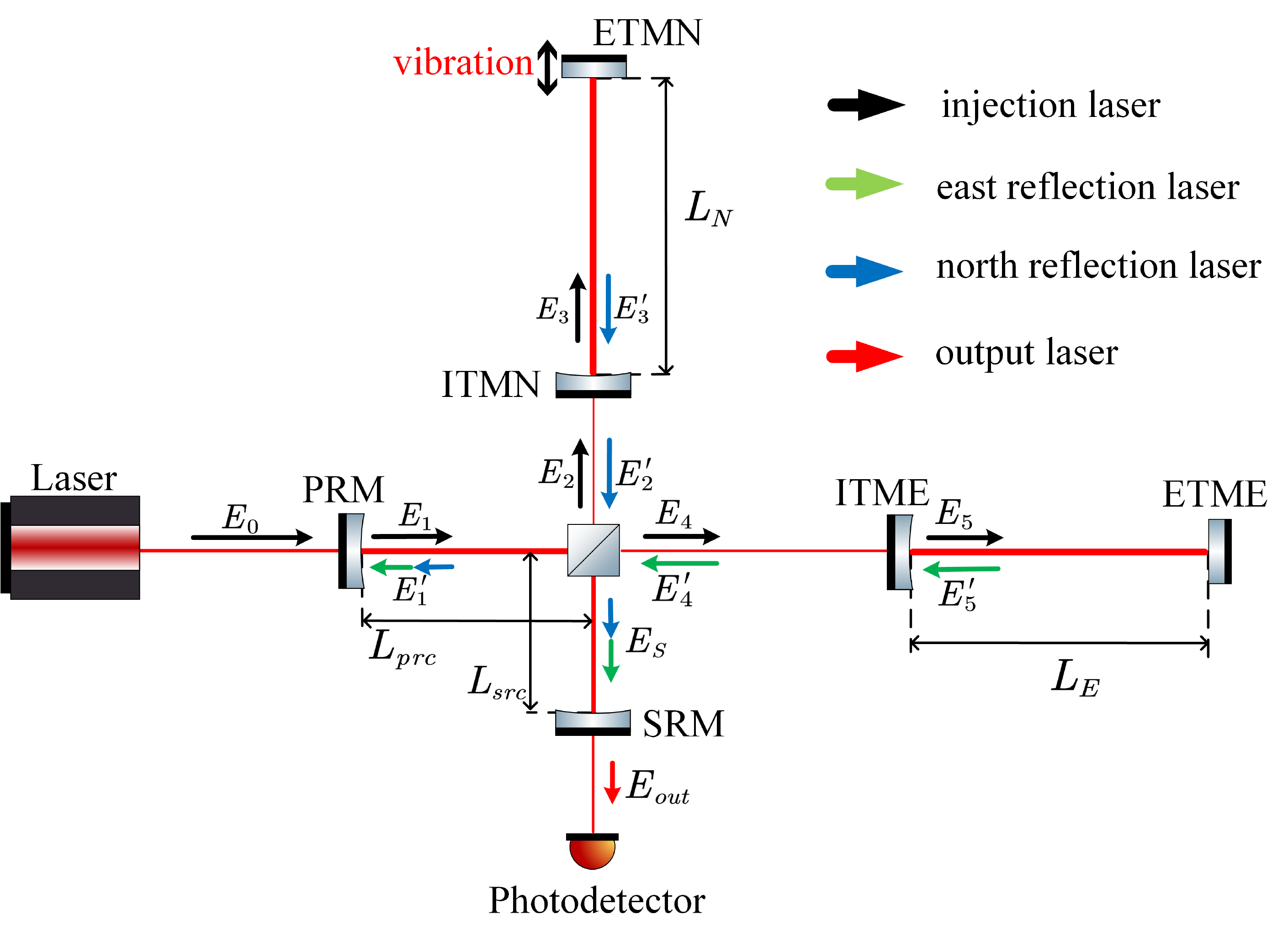}
	\caption{Diagram of dual-recycled Fabry–Perot–Michelson interferometer.}
	\label{fig.optical_model_with_cavity}
\end{figure}

We can utilise one mirror in front of splitter and photodetector to increase the light power inside the interferometer and obtain higher sensitivity \cite{freise2010interferometer}. As shown in Figure \ref{fig.optical_model_with_cavity}, the mirror PRM and splitter form a power cycle cavity and the mirror SRM and splitter form a signal cycle cavity. This configuration is most commonly called dual-recycled Fabry–Perot–Michelson interferometer. The arrows show the propogation of light fields, from the injection laser to east mirror reflection laser, north mirror reflection laser and output laser sequentially. 

\begin{theorem}\label{theorem_output_current_with_cavity}
	Assuming that the injection light power is $P_0$ and photoelectric responsivity of the photodetector is $\alpha (A/W)$, the output current of the dual-recycled Fabry–Perot–Michelson interferometer shown in Figure \ref{fig.optical_model_with_cavity} is given by
	\begin{equation}
	\begin{aligned}
	I_{out} & = \frac{\alpha}{2} G_{PRM}^2 G_{SRM}^2 P_0 \left|\left(1-\frac{t_3^2 e^{-i 2 k_0 L_E}}{1 - e^{-i 2 k_0 L_E} }\right)\right.\\
	& \quad  + \left.\left(1-\frac{t_1^2 e^{-i 2 k_0 L_N}}{e^{-i 2 k_0 x_m}- e^{-i 2 k_0 L_N} }\right)\right|^2,\\
	\end{aligned}
	\end{equation}
	where $G_{PRM}$ and $G_{SRM}$ are light power gain and signal power gain of power cycling cavity and signal cycling cavity, respectively; $t_1$, $t_3$ are transmission coefficient of mirror $ITMN$ and $ITME$, respectively.  
\end{theorem}

\begin{IEEEproof}
	Please refer to Appendix \ref{output_current_with_cavity}.
\end{IEEEproof}

\subsection{Transfer Function of Piezo-Opto-Electro-Mecahnical System in the Frequency Domain}\label{tf}
In general, the internal energy of a piezomechanical system is given by
\begin{equation}
U_{pe}=\frac{1}{2}\int{d}v\left( \mathbf{T}\cdot \mathbf{d}^T\cdot \mathbf{E}+\mathbf{E}\cdot \mathbf{d}\cdot \mathbf{T} \right),
\end{equation}
where $\mathbf{T}$ and $\mathbf{S}$ are the stress and strain fields, respectively; and  $\mathbf{E}$ and $\mathbf{D}$ are the electric field and electric displacement vector, respectively \cite{zou2016cavity}.

For one-dimension condition, the total energy of piezoelectric oscillator is given by
\begin{equation}
	\begin{aligned}
	U&=\frac{1}{2}\int{d}v\left[ T_z\cdot \left( s_{33}  T_z+d_{33} E_z \right) +E_z\cdot \left( d_{33}T_z+\varepsilon_{33} E_z \right) \right]\\
	&=\frac{1}{2} \int{dv\left(s_{33} T_z^2 \right)} + \frac{1}{2}\int{dv\left( \varepsilon_{33} E_z^2 \right)}+\\
	& \quad \frac{1}{2}\int{dv\left( T_z\cdot d_{33} E_z+E_z\cdot d_{33} T_z \right)}\\
	&\triangleq U_p+U_e+U_{pe},\\
	\end{aligned}
\end{equation}
where $U_p$, $U_e$ and $U_{pe}$ represent mechanical energy, electrical energy and electromechanical coupling energy, respectively. According to the three parts of energy, the Hamiltonian form is given by,
\begin{equation}
	\begin{aligned}
	H &=\frac{p^2}{2m}+\frac{m\omega _{M}^{2}x^2}{2}+\frac{\phi ^2}{2L_0}+\frac{q^2}{2C\left( x \right)} + \\
	  &\quad 2g\sqrt{\frac{m\omega _M}{C\left( x \right) \omega _{LC}}}xq-qV,
	\end{aligned}
\end{equation}
where $x$ and $p$ are the displacement and momentum, respectively; $q$ and $\phi$ are the charge and flux, respectively; and $g$ is the electromechanical coupling strength \cite{zou2016cavity}. We have the following results on the transfer function of the proposed piezo-opto-electro-mechanical system.
\begin{theorem}\label{theorem_transfer_function}
	Assuming that $S_{VV}^{input}$, $S_{xx}$ are signal power densities of electric signal and displacement signal of film, respectively, the transfer function of the proposed piezo-opto-electro-mechanical system in the frequency domain is given by
	\begin{equation}
		\begin{aligned}
		T(\omega) = \frac{S_{xx}(\omega)}{S_{VV}^{input}(\omega)} = |G \chi_m^{eff}(\omega)\chi_{LC}(\omega)|^2.
		\end{aligned}
	\end{equation}
	where $G$, $\chi_m^{eff}$ and $\chi_{LC}$ are coupling coefficient, effective film susceptibility and circuit's susceptibility, respectively.
\end{theorem}
\begin{IEEEproof}
	Please refer to Appendix \ref{transfer_function}.
\end{IEEEproof}

Considering the noises from circuit, film and optiocal module, the output signal of POEMS is obtained as follows,
\begin{equation}\label{Eq17}
S_{\varphi \varphi}^{output} = (2k)^2 \left(| G \chi_m^{eff} \chi_{LC}|^2 S_{VV}^{input} + S_{xx}^{th}\right) + S_{\varphi \varphi}^{imp},
\end{equation}
where $S_{VV}^{input}$, $S_{xx}^{th}$ and $S_{\varphi \varphi}^{imp}$ are the power spectral densities of input signal, film harmonic noise and optical phase noise, respectively \cite{bagci2014optical}.

\subsection{Noise Characterzation}\label{subsection.Noise Characterzation}
The main noise sources are from circuits, thermal environment and optical module. Firstly, we analyze the electrical noise of the resonant circuit and the dynamic displacement disturbance of film. Secondly, when the classical noise in the system is small enough, the influence of quantum fluctuation noise of the optical module becomes critical. Here we consider two types of noise sources: optical radiation pressure and quantum fluctuation of light. Finally, combined with the transfer function, the equivalent input noise in electrical domain can be obtained.

\subsubsection{Electrical Input Noise}
For LC resonant circuit, inductor thermal noise, shot noise and pink noise are three main types of noises. At low frequencies, the pink noise is several times larger than the shot noise, but our system works in a high frequency and can effectively reduce the influence of pink noise. Besides, since the shot noise is weaker than thermal noise under consideration, we only consider thermal noise. Additional noise due to nonstationary radio frequency interference from celluar phones, vehicles, etc., needs to be considered, but in this analysis, we will concentrate on natural sources only. The total input noise power sepectral density can be given by
\begin{equation}
N_{VV}^{electro}\left(\omega\right) = 2 k_B R_L T,
\end{equation}
where $T$ is the enviroment temperature; $R_{L}$ is the inductor equivalent resistance and $k_B$ is the Boltzmann constant.

\subsubsection{Film Thermal Noise}\label{film thermal noise}
In the LC resonant circuit, the piezoelectric film acts as a capacitor, whose noise is considered in film thermal noise. Such noise is not white in the frequency domain, while the disturbing force from environment is Gaussian white noise instead. We called this dynamic displacement disturbance of film as its thermal noise, which is driven by a disturbing force from environment.
\begin{theorem}\label{theorem_film_thermal_noise}
	In equilibrium, the environment exerts a disturbing force $\eta$ on the damped harmonic oscillator. According to the white noise hypothesis, $\eta \left( t \right)$ satisfy $\langle \eta \left( t \right) \rangle =0$ and $\langle \eta \left( t \right) \eta \left( \tau \right) \rangle =2\alpha \delta \left( t-\tau \right)$. The equivalent harmonic noise power spectrum of the piezoelectric film is given by
	\begin{equation}
		\begin{aligned}
		N_{xx}^{film}\left( \omega \right) &=\ 2\alpha _{ex}|\chi _m\left( \omega \right) |^2,\\
		\end{aligned}
	\end{equation}
	where $\alpha_{ex}=\alpha m_{eff}^{2}$ represents the strength of noise. 
\end{theorem}	
\begin{IEEEproof}
	Please refer to Appendix \ref{film_thermal_noise}.
\end{IEEEproof}

\subsubsection{Optical Module Noise}\label{optical module noise}
In the optical cavity, the equivalent dynamic displacement noises of the film, caused by quantum noise and optical radiation pressure noise, are given by
\begin{equation}
	N_{xx}^{\text{imp}}\left( \omega \right) =\frac{\kappa}{16\overline{n}_{cav}G_{opt}^2}\left( 1+4\frac{\omega ^2}{\kappa ^2} \right),
\end{equation}
and
\begin{equation}
	N_{xx}^{FF}\left( \omega \right)
	= \overline{n}_{\text{cav}} \frac{4 \hbar ^2 G_{opt}^2}{\kappa} \left (1+4\frac{\omega ^2}{\kappa ^2 }\right )^{-1} |\chi_m(\omega)|^ 2,
\end{equation}
respectively, where $\kappa$ and $G$ are cavity decay and coupling coefficient, respectively; and $\overline{n}_{cav}$ is the average number of photons in cavity \cite{aspelmeyer2014cavity}.

The optical module noise under consideration can be given by
\begin{equation}
	\begin{aligned}
		N_{xx}^{optical} \left(\omega \right)
		&=N_{xx}^{imp}\left( \omega \right) + N_{xx}^{FF}\left( \omega \right) \\
		&=\frac{C}{16 \overline{n}_{cav}} + \frac{4\hbar ^2 \overline{n}_{cav}}{C} {\left|\chi _m\left( \omega \right) \right|}^2\\
		&\ge2 \sqrt{\frac{C}{16 \overline{n}_{cav}} \cdot  \frac{4\hbar ^2 \overline{n}_{cav}}{C} {\left|\chi _m\left( \omega \right) \right|}^2}\\
		&=\hbar\left|\chi _m \left(\omega \right) \right|,\\
	\end{aligned}
\end{equation}
where $C \triangleq \kappa \left( 1+4\frac{\omega ^2}{\kappa ^2} \right) /G_{opt}^2$ is a defined coefficient. It should to be noted that the noise of optical module is typically represented by the phase uncertainty. Since the optical module is employed to measure the film vibration, the corresponding phase noise is as follows,
\begin{equation}
N_{\varphi \varphi}^{optical}\left( \omega \right) = (2k)^2 N_{xx}^{optical}\left( \omega \right),
\end{equation}
where $k$ is the magnitude of the light wave vector.

\subsubsection{Output Noise}
Considering the main noise components in the system, the output noise power spectral density at resonance frequency $\omega_r$ is given by
\begin{equation}\label{Eq24}
N_{\varphi\varphi}^{output} = (2k)^2\left( \left| G \chi _{m}^{eff} \chi _{LC} \right|^2 N_{VV}^{electro}+ N_{xx}^{film} \right) + N_{\varphi\varphi}^{optical},
\end{equation}
where $\chi_{LC}$, $\chi_m^{eff}$ and $G$ are intoduced in Theorem \ref{theorem_transfer_function}. 

\subsection{Sensitivity Limit}
According to Eq.\ref{Eq17} and Eq.\ref{Eq24}, the system output signal to noise ratio can be given by
\begin{equation}
	\text{SNR} = \frac{(2k)^2 S_{VV}^{input}\left|G \chi _{m}^{eff} \chi _{LC}\right|^2}{(2k)^2\left( \left| G \chi _{m}^{eff} \chi _{LC} \right|^2 N_{VV}^{electro}+ N_{xx}^{film} \right) + N_{\varphi\varphi}^{optical}}.
\end{equation}

The reference sensitivity power level of evolved universal terrestrial radio access base station is measured under a
throughput requirement for a specified reference measurement channel. The modulation method is QPSK and code rate is $1/3$ \cite{etsi2019136}, leading to symbol-level SNR about $0$ dB. For the proposed system, the minimum signal power spectral density for SNR$=0$ dB is given by
\begin{equation}
	S_{VV}^{min} = \underset{S_{VV}^{input}}{\text{Arg}}\left\lbrace \text{SNR}=0\ \text{dB}\right\rbrace .
\end{equation}
Assuming that the signal bandwdth is $B$ and input impedance is $R_i$, the minimum signal power is given by
\begin{equation}
	P_{min} = \frac{S_{VV}^{min} B}{R_i}.
\end{equation}

\section{Numerical Results on the Components}

\subsection{Piezoelectric Film}
\begin{figure}[!t]
	\centering
	\includegraphics[width = 3.5in]{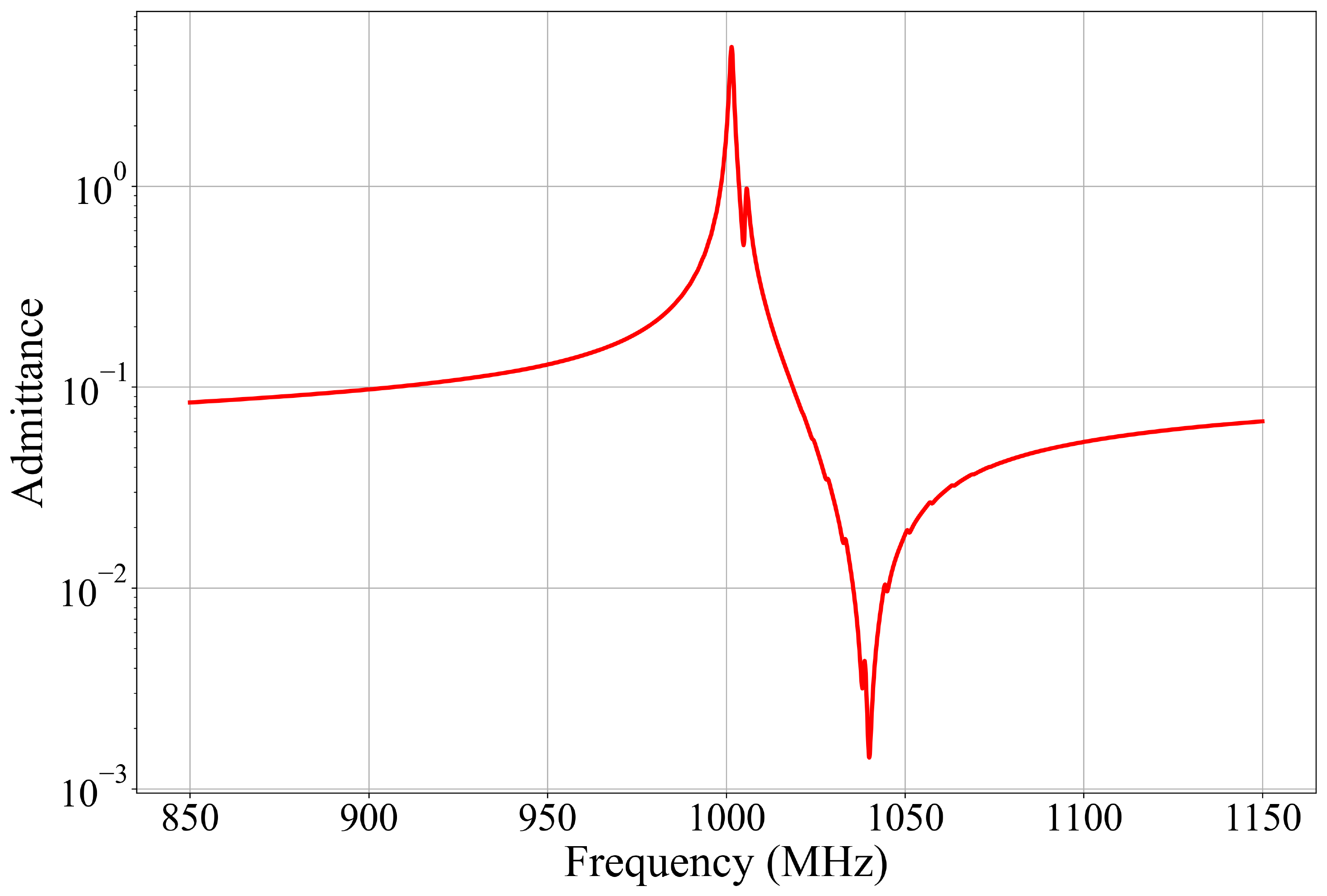}
	\caption{The simulation of piezoelectric film admittance curves.}
	\label{fig.BVD_curve}
\end{figure}

COMSOL Multiphysics${\circledR}$ is a large-scale advanced numerical simulation software. To simulate all kinds of physical processes, COMSOL Multiphysics${\circledR}$ is adopted to realizes highly accurate numerical simulation with high-efficiency computing performance and outstanding multi field bidirectional direct coupling analysis ability. In order to verify the rationality of sensitivity analysis, the admittance curve of the model was simulated and verified by COMSOL Multiphysics${\circledR}$ software \cite{multiphysics2018comsol}. We adopt an AlN film that is $500$ microns long, $500$ microns wide and $5.5$ microns high. The sweep frequency range is set to be $0.85$ GHz to $1.15$ GHz to include the film resonance frequency. In order to save the running time of the program, the peak response location can be estimated through the theoretical calculation to guide the setting of the sweep frequency range. Finally, the admittance curves is shown in Figure \ref{fig.BVD_curve}. The peak response occurs at the position where frequency approximately equals $1001$ MHz.

%
%
%

\begin{figure}[!t]
	\centering
	\includegraphics[width = 3.5in]{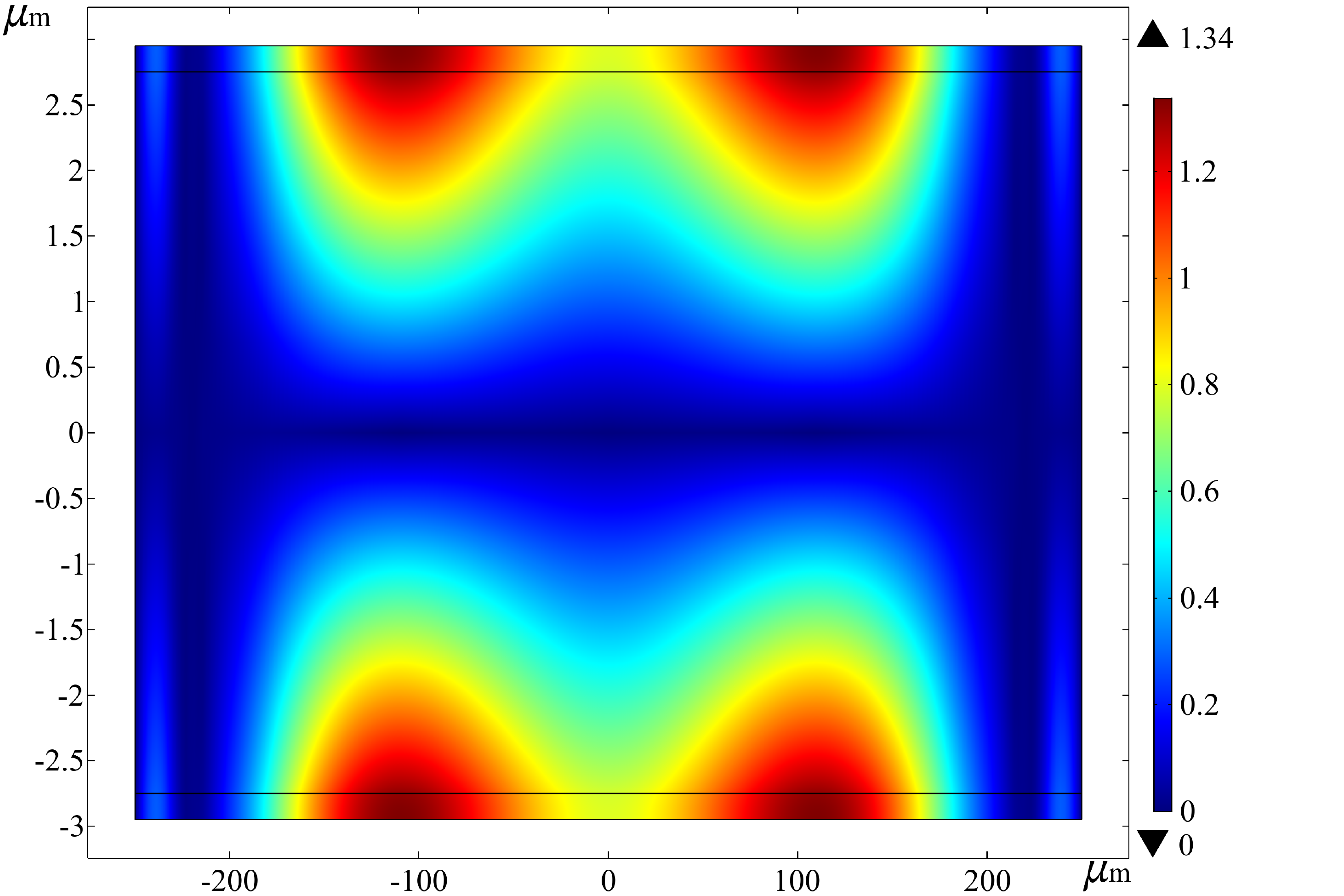}
	\caption{A regional 2D vibaration schematic of piezoelectric film under the RF signal excitation.}
	\label{fig.COMSOL_vibration}
\end{figure}

Under the same conditions, the excitation signal frequency is set to be resonance frequency, where the maximum admittance is obtained. Assuming that the amplitude of sinusoidal signal is $1$ V with zero offset, the film variation along z-axis is shown in Figure \ref{fig.COMSOL_vibration}. Compared with the vibration equations in Section \ref{Response to excitation}, the theoretical calculation results can well match the simulation results. The relevant parameters in the simulation can be seen in Table \ref{Parameters used in COMSOL} \cite{auld1973acoustic}. According to Eq. \ref{Eq7}, the theoretical amplitude of surface displacement is about $1.5$ nm. The maximum amplitude shown in Figure \ref{fig.COMSOL_vibration} is about $1.34$ nm, matching the theoretical value within $0.6$ dB with the same order of magnitude. The possible reason is the superposition of other vibration modes.

\begin{table}[htbp]
	\centering
	\caption{Typical Parameters used in COMSOL simulation}
	\label{Parameters used in COMSOL}
	\begin{tabular}{|l|c|c|}  
		\hline
		& & \\[-6pt]
		Name & Symbol & Value \\ 
		\hline
		& & \\[-6pt]
		piezoelectric constant & $d_{33}$ & $4.98\times 10^{-12}$ C/N \\ 
		\hline
		& & \\[-6pt]
		piezoelectric constant & $e_{33}$ & $1.55$ C/m$^2$ \\ 
		\hline
		& & \\[-6pt]
		piezoelectric constant & $s_{33}$ & $3.21\times 10^{-12}$ m$^2$/N \\ 
		\hline
		& & \\[-6pt]
		modulus of elasticity & $c^D$ & $4.19 \times 10^{11}$ N/m$^2$ \\ 
		\hline
		& & \\[-6pt]
		permittivity constant& $\varepsilon^{S}$ & $7.97\times 10^{-11}$ F/m \\ 
		\hline
		& & \\[-6pt]
		density of film & $\rho$ & $3300$ kg/m$^3$ \\ 
		\hline
		& & \\[-6pt]
		piezoelectric film thickness & $L_T$ & $5.5$ um \\ 
		\hline
		& & \\[-6pt]
		piezoelectric film width & $W$ & $500$ um \\ 
		\hline
		& & \\[-6pt]
		piezoelectric film length & $L$ & $500$ um \\ 
		\hline
		& & \\[-6pt]
		quality factor of film & $Q$ & $1000$ \\ 
		\hline
		& & \\[-6pt]
		resonant frequency of circuit & $f_{LC}$ & $1$ GHz \\ 
		\hline
		& & \\[-6pt]
		resonant frequency of film & $f_{m}$ & $1$ GHz \\ 
		\hline
	\end{tabular}
\end{table}

\subsection{Optical Detection Module}
Frequency domain INterferomEter Simulation SoftwarE (FINESSE) is a simulation program for interferometers. For a given optical setup, it computes the light field amplitudes at every point in the interferometer assuming a steady state. The interferometer description is translated into a set of linear equations that are solved numerically, where extensive analysis on the performance prediction be performed, including computing the modulation-demodulation error signals and transfer functions. It can also perform the analysis using plane waves or Hermite-Gauss modes, while the latter one allows computing the effects of mode matching and misalignments. In addition, the error signals for automatic alignment systems can be simulated \cite{finesse}. We use FINESSE to perform simulation. The parameter settings in the simulation are partly from the reference example \cite{freise2010interferometer}, while the arm length and components tuning are optimized to obtain peak sensitivity at $1$ GHz. Such parameters can provide guidelines for the real fabrication. The simulation includes two parts: response to mirror vibration signal and noise limited sensitivity of optical module. Detailed parameters can be seen in Table \ref{Parameters used in FINESSE}.

\subsubsection{Response to Mirror Vibration}
\begin{figure}[!t]
	\centering
	\includegraphics[width = 3.3in]{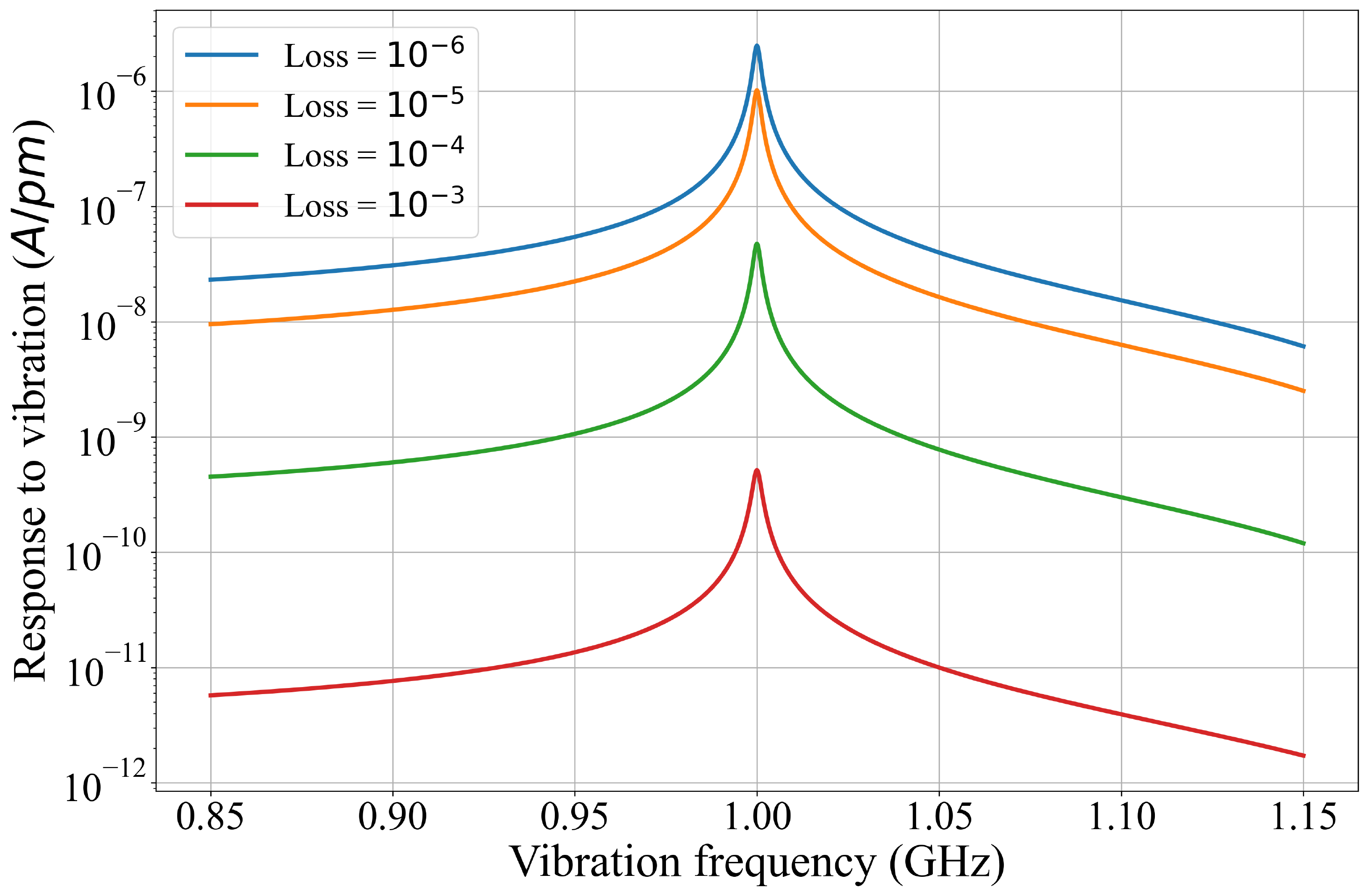}
	\caption{The response to mirror vibration with respect to carrier frequency.}
	\label{fig.response_to_mirror_vibration_FINESSE_A}
\end{figure}

This is a simple FINESSE simulation showing how the response signal can be modulated by the vibration signal on the end mirror. Assume that the north end mirror position in Figure 5 is modulated by a periodic signal $x_m = a_s \text{cos}\left( \omega_s t + \varphi_s\right)$, the output light power varies with amplitude $a_s$ and frequency $\omega_s$. In the simulation, the frequencies varies from $0.85$ GHz to $1.15$ GHz, as shown in Figure \ref{fig.response_to_mirror_vibration_FINESSE_A}. Besides, the optical components parameters can significantlly change the response characteristics. Specfic parameters has been shown in Table \ref{Parameters used in FINESSE}. The maximal gain occurs at the resonant frequecy $1$ GHz, while the responsivity decreases with the frequency shift. Obviously, the response signal amplitude and power loss coefficient are negatively correlated. We scale the response signal amplitude to photocurrent with unit $\text{A/{pm}}$. The scaling factor (from Watts to Amperes) is given by
\begin{equation}
C_{ampere}=\frac{e\lambda_0}{hc} (\text{A/W}),
\end{equation}
where $e$, $h$, $\lambda_0$ and $c$ are the electron charge, the Planck’s constant, the speed of light and the laser wavelength, respectively.

\subsubsection{Noise Limit Sensitivity of Optical Module}
\begin{figure}[!t]
	\centering
	\includegraphics[width = 3.5in]{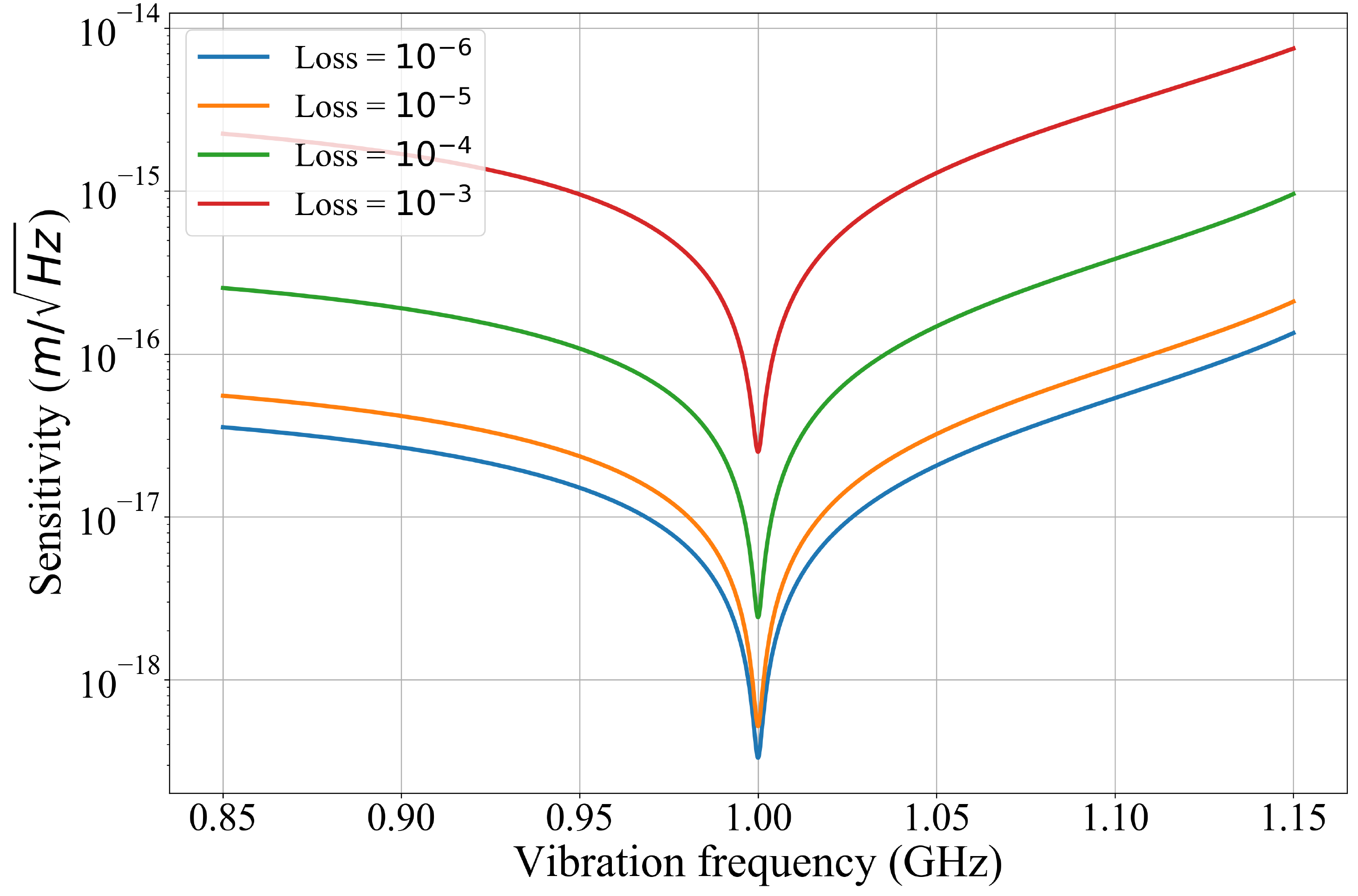}
	\caption{The noise limited sensitivity.}
	\label{fig.noise_limited_sensitivity_FINESSE}
\end{figure}

Shot noise is a type of readout noise in experimental observation. When the number of energy-carrying particles (such as electrons in a circuit or photons in an optical instrument) in the observation is small enough to cause observable statistical fluctuations in data reading, the statistical readout fluctuations are called shot noise. Uncertainty of quantum noise and optical radiation pressure noise is described in Section \ref{optical module noise}. It is reported that the noise limited sensitivity at low frequencies (a few tens of Hertz) can reach up to $10^{-23} \text{m}/\sqrt{\text{Hz}}$, which size is several kilometers \cite{freise2010interferometer}. This simulation shows the noise limited sensitivity of the proposed optical module at higher frequencies and smaller size. As shown in Figure \ref{fig.noise_limited_sensitivity_FINESSE}, the highest sensitivity occurs at the resonant frequecy $1$ GHz. Similarly, the detection sensitivity and power loss coefficient are negatively correlated.

\begin{table}[htbp]
	\centering
	\caption{Parameters used in FINESSE simulation}
	\label{Parameters used in FINESSE}
	\begin{tabular}{|l|c|c|}  
		\hline
		& \\[-6pt]
		Parameter & Value \\
		\hline
		& \\[-6pt]
		east end mirror transmissivity  &  $5\times 10^{-6}$\\
		\hline
		& \\[-6pt]
		east end mirror power loss  & $1\times 10^{-5}$ \\
		\hline
		& \\[-6pt]
		east incident mirror transmissivity  &  $0.014$\\
		\hline
		& \\[-6pt]
		east incident mirror loss  & $1\times 10^{-5}$ \\
		\hline
		& \\ [-6pt]
		north end mirror transmissivity &  $5\times 10^{-6}$\\
		\hline
		& \\[-6pt]
		north end mirror power loss & $1\times 10^{-5}$ \\
		\hline
		& \\ [-6pt]
		north incident mirror transmissivity &  $0.014$\\
		\hline
		& \\[-6pt]
		north incident mirror power loss & $1\times 10^{-5}$ \\
		\hline
		& \\[-6pt]
		beam splitter mirror transmissivity & 0.5 \\
		\hline
		& \\[-6pt]
		beam splitter mirror reflectivity & 0.5 \\
		\hline
		& \\[-6pt]
		north arm length & 7.5 cm \\
		\hline
		& \\[-6pt]
		east arm length & 7.5 cm \\
		\hline
		& \\[-6pt]
		laser power & 1 W \\
		\hline
		& \\[-6pt]
		wavelength & 1064 nm\\
		\hline
	\end{tabular}
\end{table}

\section{System Performance Evaluation}
\subsection{Gain from Low Noise Amplifier}
\begin{figure}[!t]
	\centering
	\includegraphics[width = 3.0in]{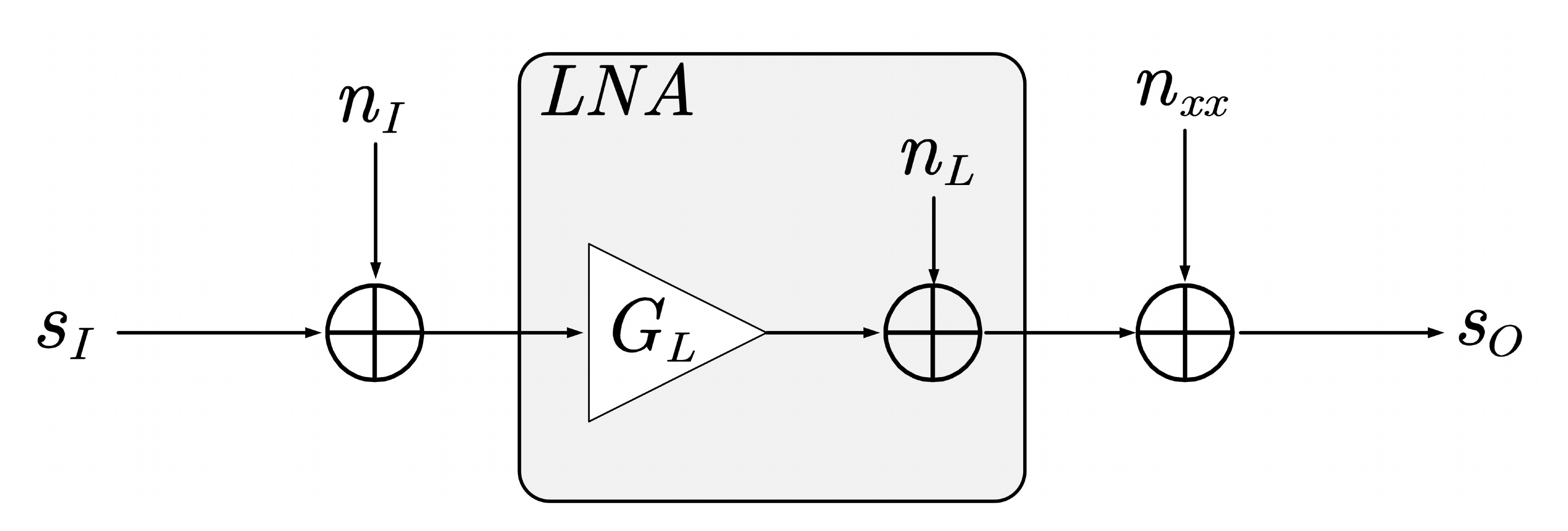}
	\caption{The signal flow graph of the receiver system with an LNA.}
	\label{fig.LNA}
\end{figure}

We evaluate the link performance via employing  a low noise amplifier (LNA) in the system. Assume that the input signal power spectral density is $s_I$. The next level has certain noise $n_{xx}$. The system SNR without LNA is given by
\begin{equation}
\text{SNR}_f=\frac{s_I}{n_I+n_{xx}}.
\end{equation}
Then, we introduce an LNA with noise $n_L$ and gain $G_L>1$, as shown in Figure \ref{fig.LNA}. The system SNR at the LNA ouput can be given by
\begin{equation}
\text{SNR}_{f}^{L}=\frac{G_L s_I}{G_L n_I + n_L + n_{xx}}.
\end{equation}
The SNR gain from LNA is given as follows
\begin{equation}
\text{SNR}_{f}^{L}-\text{SNR}_f=\frac{s_I\left[ \left( G_L-1 \right) n_{xx}-n_L \right]}{\left( G_L n_I + n_L+n_{xx} \right) \left( n_I + n_{xx} \right)}.
\end{equation}
When $G_L>G_{L}^{*}=n_L/n_{xx}+1$, the LNA will improve the SNR of the system.

\subsection{System Sensitivity}\label{subsection.system sensitivity}
\begin{figure*}[!t]
	\centering
	\includegraphics[width = 4.5in]{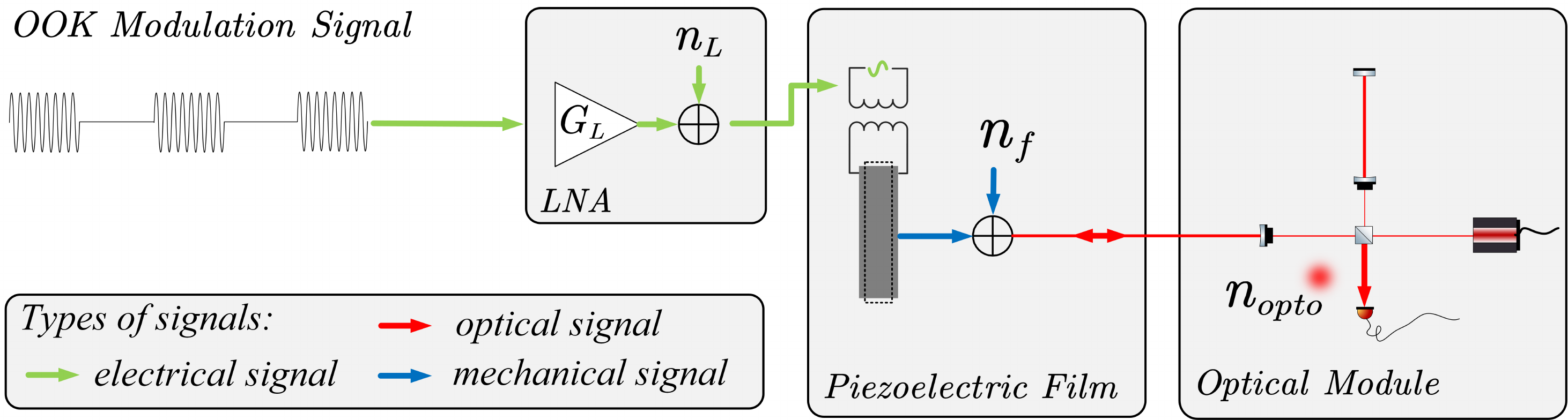}
	\caption{The signal flow graph in the system link simulation.}
	\label{fig.full_link_simulation_diagram}
\end{figure*}

The system link simulation diagram is shown in Figure \ref{fig.full_link_simulation_diagram}. The OOK modulation signal is transmitted to the resonant circuit, consisted of inductor and piezoelectric film. Then, the RF signal excites the piezoelectric film to vibrate, leading the variation of light phase in the optical module. Finally, the photodetector transfers the light signal to electric signal for symbol detection. Detailed parameters used in the simulation are shown in Table \ref{Parameters used in link}.

\begin{table}[htbp]
	\centering
	\caption{Typical Parameters used in system link simulation}
	\label{Parameters used in link}
	\begin{tabular}{|l|c|c|}
		\hline
		& & \\[-6pt]
		Name & Symbol & Value \\ 
		\hline
		& & \\[-6pt]
		enviroment temperature & $T$ & $300$ K \\ 
		\hline
		& & \\[-6pt]
		carrier frequency & $f_c$ & $1$ GHz \\ 
		\hline
		& & \\[-6pt]
		signal banwidth & BW & $1$ kHz \\ 
		\hline
		& & \\[-6pt]
		modulation method & MM & OOK \\  
		\hline
		& & \\[-6pt]
		LNA gain & $G_L$ & $30$ dB \\ 
		\hline
		& & \\[-6pt]
		LNA noise temperature & $T_{L}$ & $25$ K \\ 
		\hline
		& & \\[-6pt]
		LC circuit noise PSD & $n_{LC}$ & $8.3\times 10^{-21}\ \text{V}^2\text{/Hz}$\\ 
		\hline
		& & \\[-6pt]
		film noise PSD & $n_f$ & $3.4\times 10^{-55}\ \text{m}^2\text{/Hz}$ \\ 
		\hline
		\multicolumn{3}{|l|}{}\\[-6pt]
		\multicolumn{3}{|l|}{Note: PSD represents power spectral density, whose units for electric and}\\
		\multicolumn{3}{|l|}{}\\[-6pt]
		\multicolumn{3}{|l|}{displacement signals are $\text{V}^2/\text{Hz}$ and $\text{m}^2/\text{Hz}$, respectively.}\\
		\hline
	\end{tabular}
\end{table}

\begin{figure*}[!t]
	\centering
	\includegraphics[width = 4.5in]{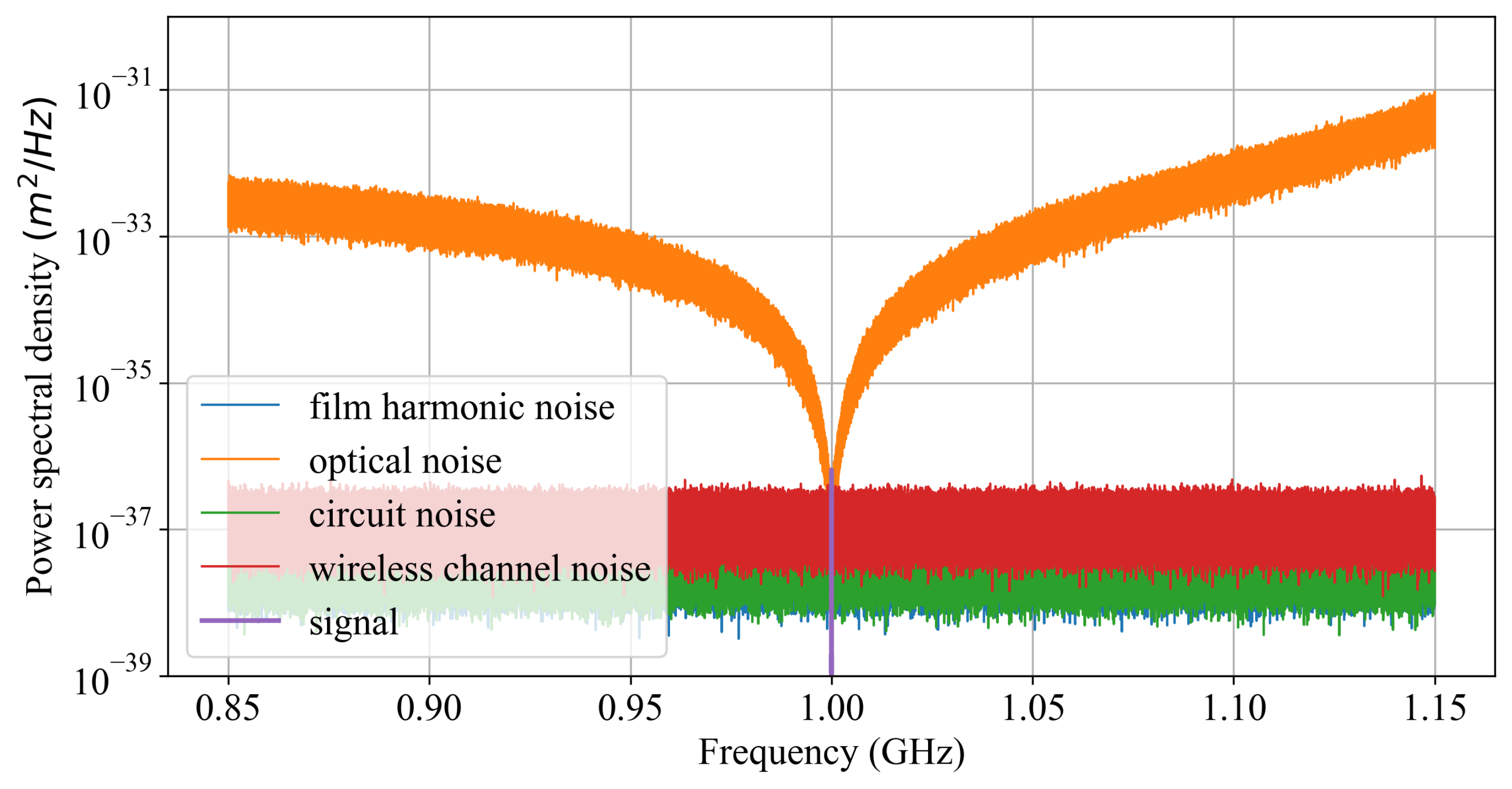}
	\caption{Signal and noises power spectral densities with input power $-160$ dBm.}
	\label{fig.full_link_simulation_result}
\end{figure*}

For the OOK modulation, the carrier frequency is set to $1$ GHz and assume the bandwidth is 1kHz. We set the LNA gain to be a typical value $G_L = 30$ dB \cite{wadefalk2005very}. The electrical signal excites the piezoelectric film to vibrate through a coupling circuit, where the response signal is characterized in Section \ref{tf}. This process also introduces equivalent film thermal noise in Section \ref{film thermal noise}, whose noise power spectral density is $n_{f}=3.4 \times 10^{-55} \text{m}^2/\text{Hz}$. Assuming that the lens power loss is $1\times 10^{-5}$, the noise characterization of the POEMS with contributions from circuit Johnson noise (green), optical noise (orange), film thermal noise (blue) and wireless channel noise (red) are shown in Figure \ref{fig.full_link_simulation_result}. It can be seen that the signal can be differentiated from the noise.

\begin{figure}[!t]
	\centering
	\includegraphics[width = 3.5 in]{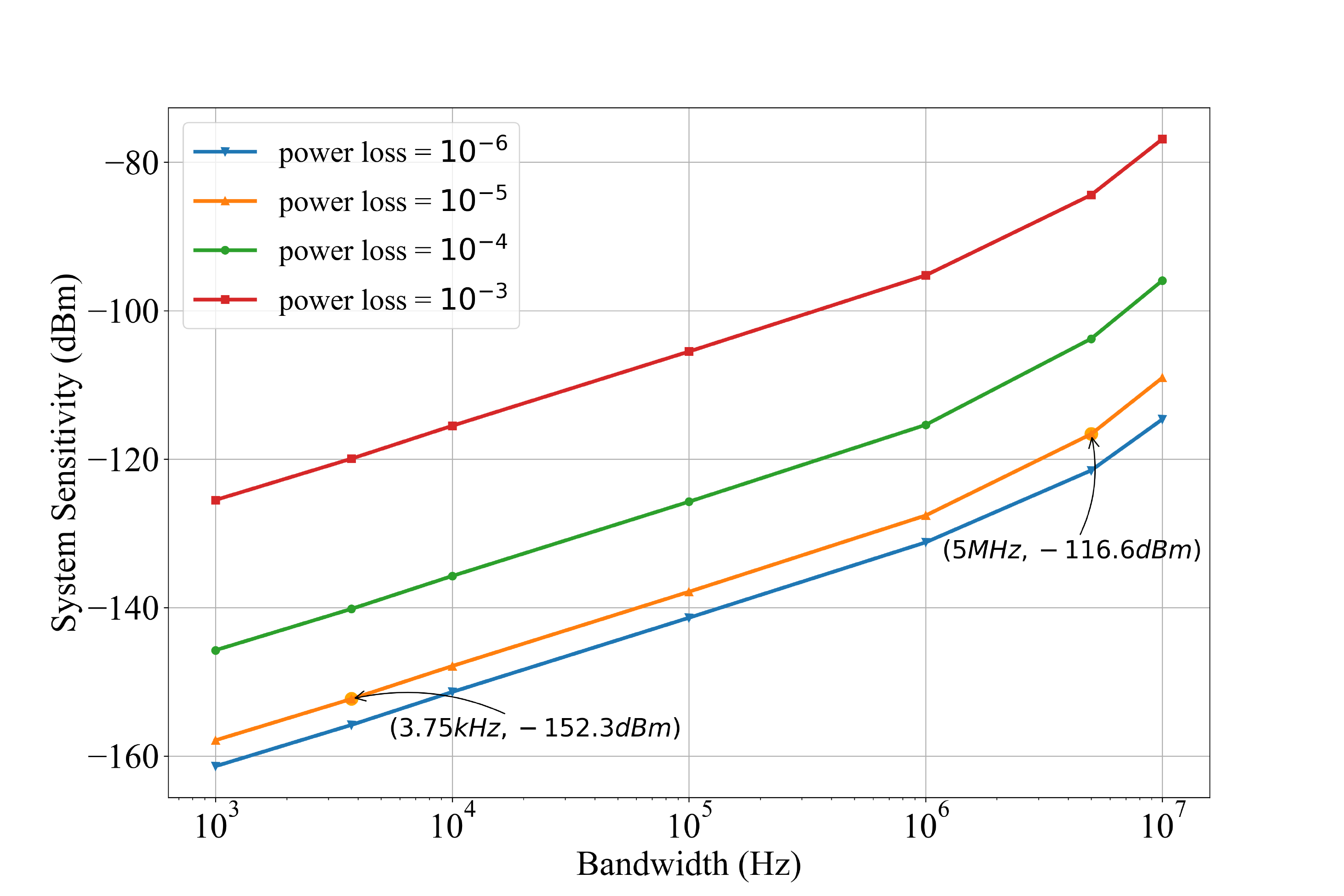}
	\caption{The system sensitivity to different bandwidths and lens power loss coefficients with LNA in consideration. The bandwidth varies from $1$ kHz to $10$ MHz.}
	\label{fig.system_sensitivity_with_LNA}
\end{figure}

\begin{figure}[!t]
	\centering
	\includegraphics[width = 3.5 in]{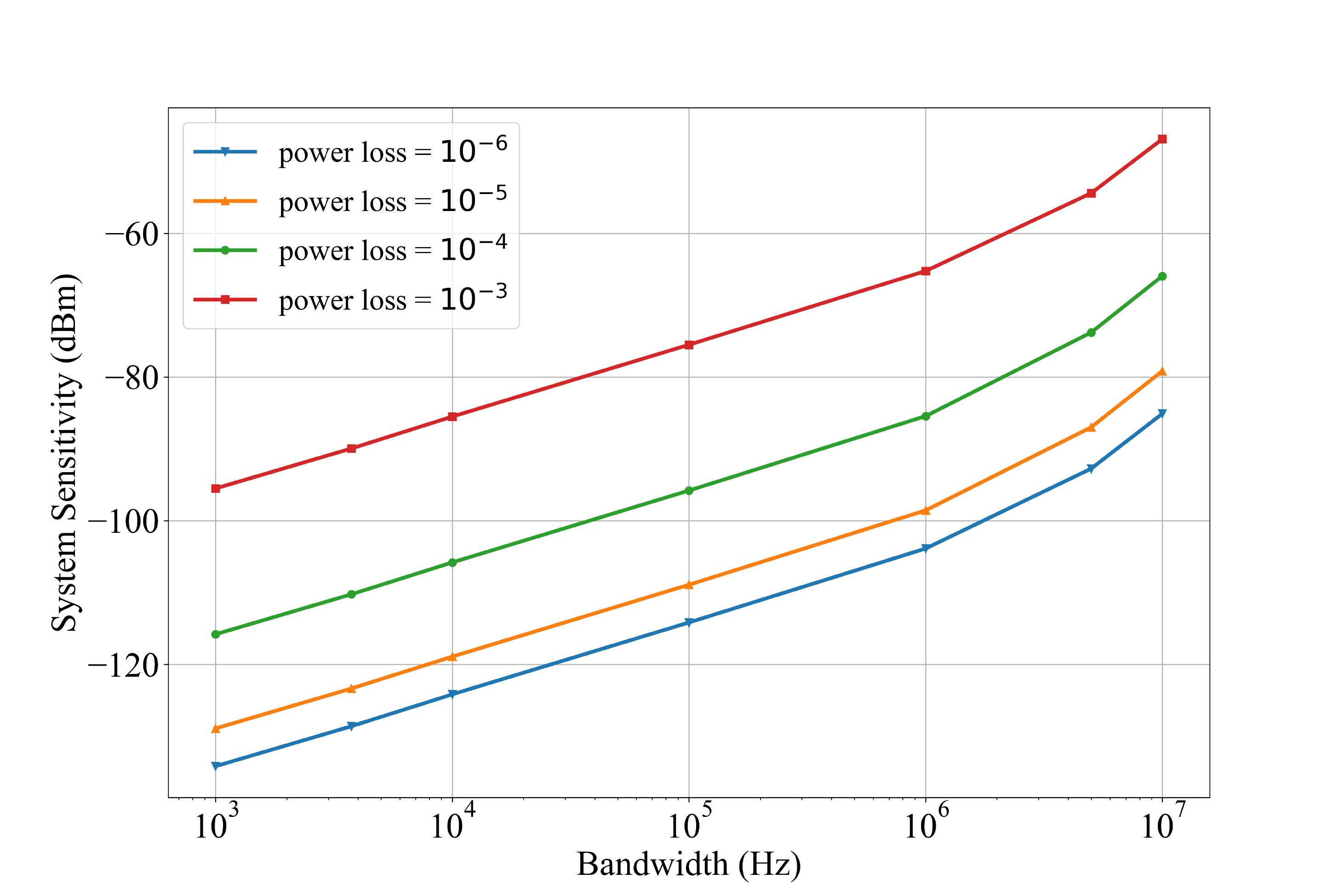}
	\caption{The system sensitivity to different bandwidths and lens power loss coefficients without LNA in consideration. The bandwidth varies from $1$ kHz to $10$ MHz.}
	\label{fig.system_sensitivity_without_LNA}
\end{figure}

According to the theoretical analysis in Section \ref{Sensitivity_Analysis}, the syetem sensitivity critically depends on the optical module detection sensitivity and channel bandwidth, while the detection sensitivity depends on the lens power loss coefficient. Figure \ref{fig.system_sensitivity_with_LNA} and Figure \ref{fig.system_sensitivity_without_LNA} show the system sensitivity variation with the signal bandwidth under different loss coefficients. Assuming that the power loss coefficient is $1\times 10^{-5}$, for system with LNA, the system sensitivity at bandwidths $3.75$ kHz and $5$ MHz are $-152.3$ dBm and $-116.6$ dBm, respectively. Lower bandwidth and lower power loss coefficient lead to higher sensitivity. 

In Appendix \ref{Reference_Sensitivity_of_Base_Station}, we summarize the reference sensitivity power levels of evolved universal terrestrial radio access (E-UTRA) base station and narrowband internet of thing (NB-IoT) base station in 4G LTE system. The highest sensitivities for NB-IoT and E-UTRA base stations are $-133.7$ dBm at bandwidth $3.75$ kHz and $-101.5$ dBm at bandwidth $5$ MHz, respectively. It shows that about 18 dB at bandwidth $3.75$ kHz and 15 dB at bandwidth $5$ MHz gain can be predicted by this prototype design with LNA. Also, we calculate the equivalent sensitivity of atom-based system in Appedix \ref{Reference_Sensitivity_of_Atom_System}, which is about $-120$ dBm at bandwidth $1$ Hz. For a fain comparison, we normalize the bandwith to be $3.75$ kHz and compare it with proposed system without LNA. The proposed prototype design can obtain a gain of $40$ dB.

\subsection{Signal to Noise Ratio, Bit Error Rate and Capacity}
\begin{figure}[!t]
	\centering
	\includegraphics[width = 3.5in]{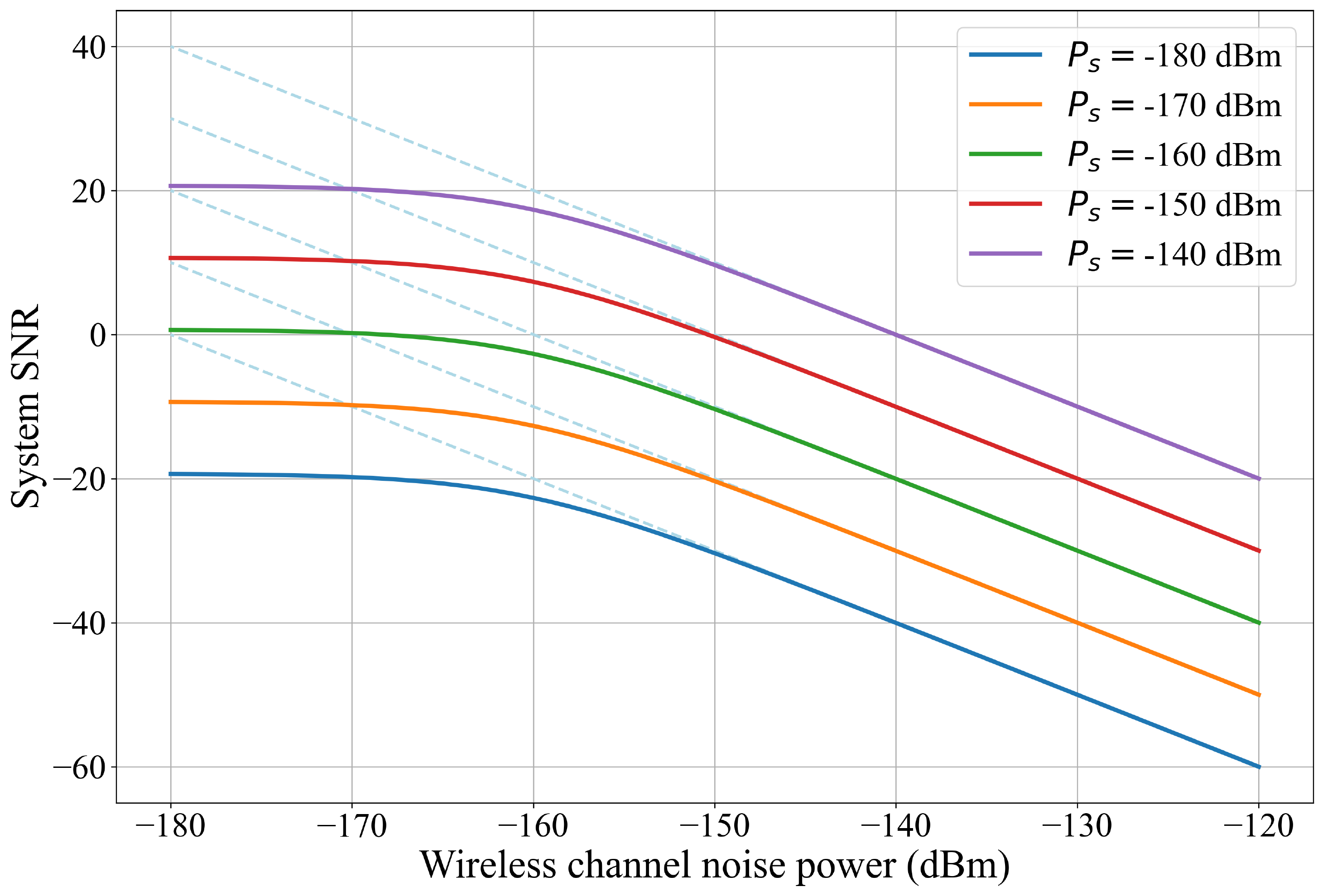}
	\caption{The SNR of the output signal from optical detection under different wireless channe noise power.}
	\label{fig.full_link_simulation_SNR}
\end{figure}

Figure \ref{fig.full_link_simulation_SNR} shows the system output SNR with external wireless channel noise power. When the signal power is $-150$ dBm and the wireless channel noise power is $-160$ dBm, more than $7$ dB SNR can be achieved.

\begin{figure}[!t]
	\centering
	\includegraphics[width = 3.5in]{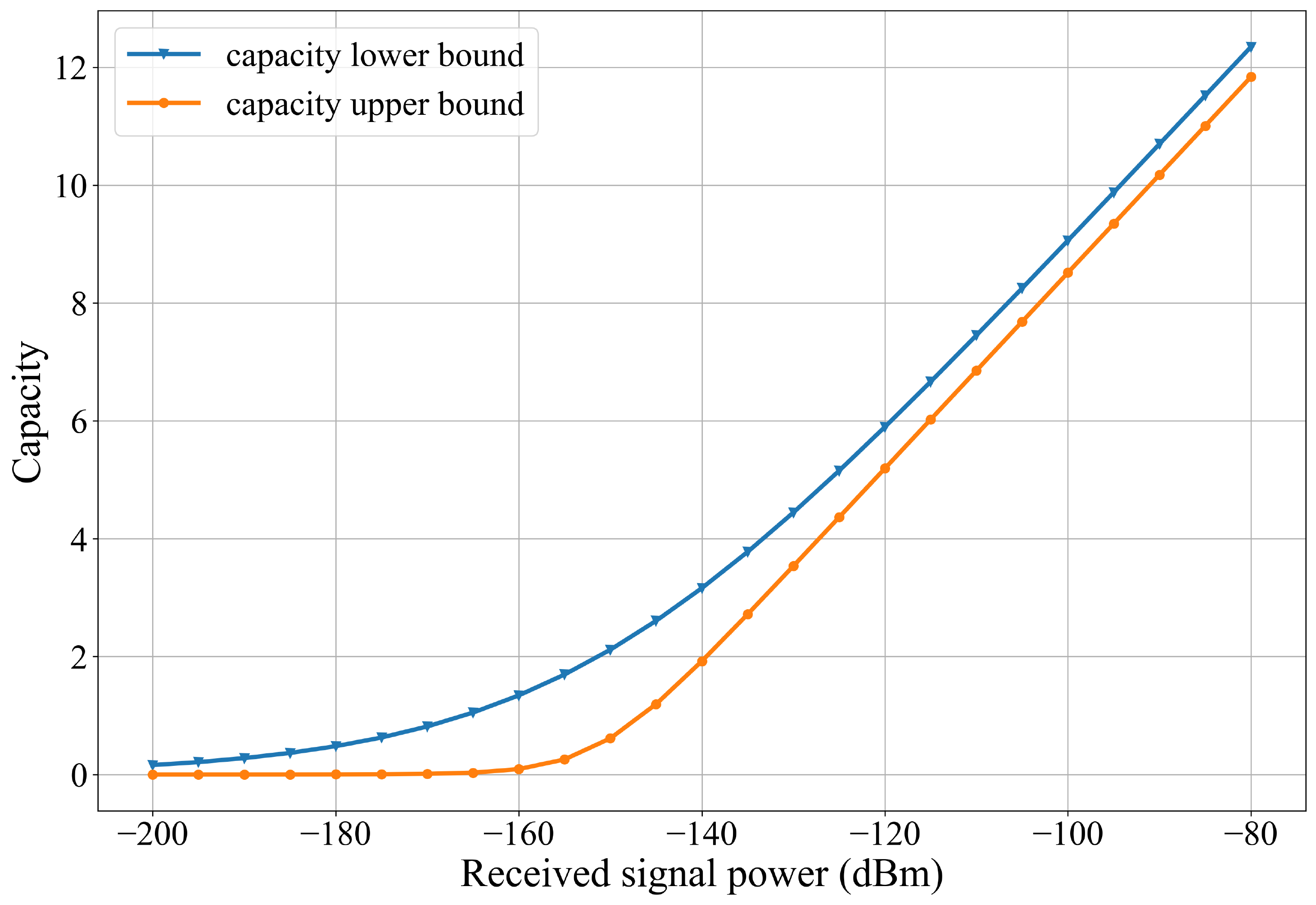}
	\caption{Upper and lower bounds on the capacity of the system under consideration.}
	\label{fig.full_link_simulation_capacity}
\end{figure}

We refer to IM/DD Gaussian channel capacity under vector modulation, to evaluate the achievable rate of the proposed piezo-opto-electro-mechanical system\cite{wang2013tight}. Assuming wireless channel noise power is $-165$ dBm, we get upper and lower bounds on the channel capacity results under different received power, as shown in Figure \ref{fig.full_link_simulation_capacity}.

\begin{figure}[!t]
	\centering
	\includegraphics[width = 3.5in]{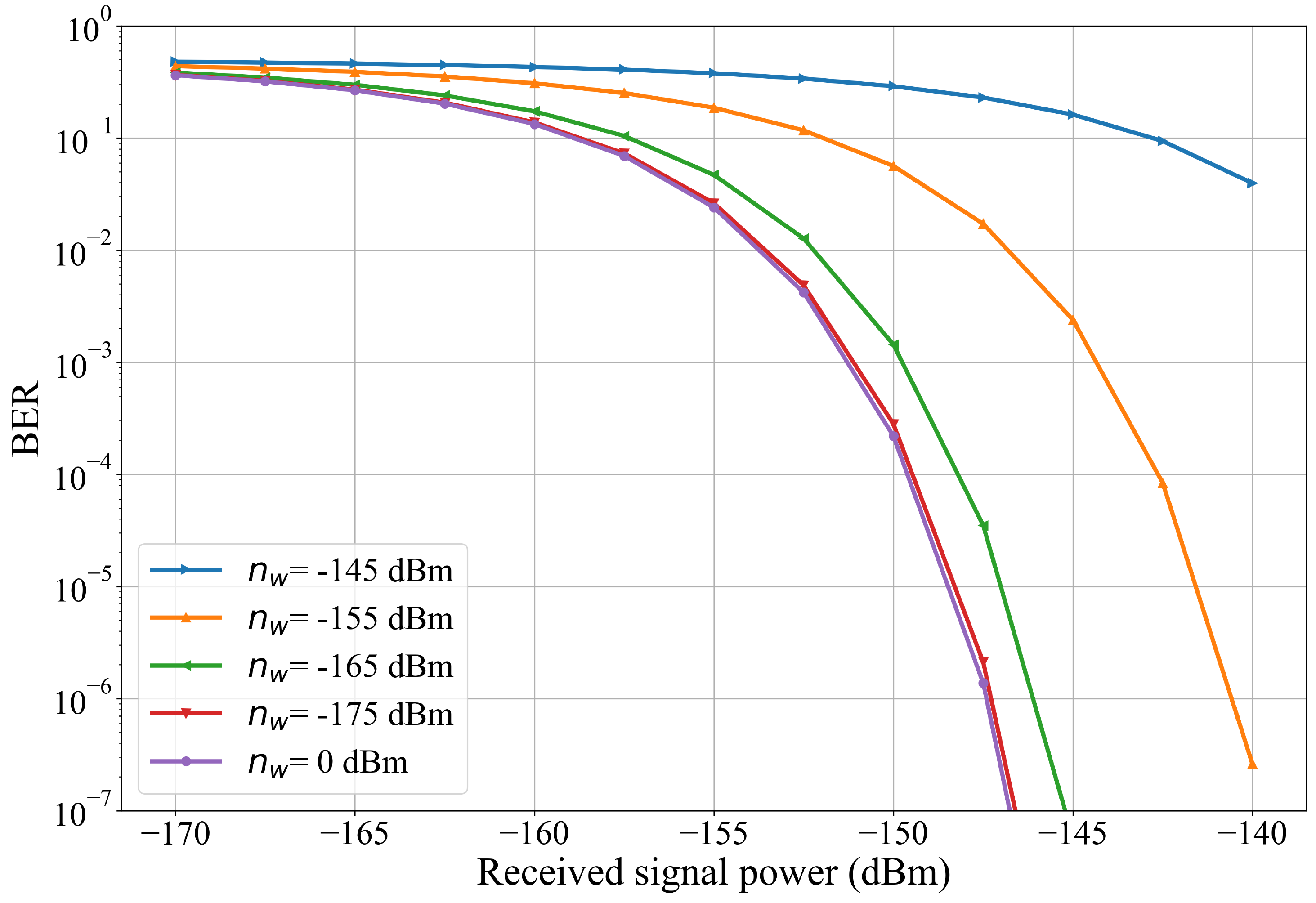}
	\caption{The bit error rate under different wireless channel noise power level.}
	\label{fig.full_link_simulation_BER}
\end{figure}

Under different input power and wireless channel noise power, we perform the bit error rate simulation of the $1$ kbps OOK signal, as shown in Figure \ref{fig.full_link_simulation_BER}. The BER is lower than $10^{-3}$ when the received power is $-150$ dBm and the wireless channel noise power is lower than $-175$ dBm.

\section{Conclusion}
We have proposed a high sensitivity piezo-opto-electro-mechanical receiver system, which utilizes optical module to detect the piezoelectric film vibration driven by RF signal. Based on the analysis of piezoelectric vibration and optical response to the vibration, we have analyzed the system transfer function in the frequency domain. Both theoretical results and numerical/simulation results are provided to test the feasibility of the proposed system. For OOK modulation signal with $5$ MHz bandwidth and $1$ GHz carrier frequency, the system receiving sensitivity can be predicted to $-118.9$ dBm, which significantly outperforms that of the reference sensitivity power of the evolved universal terrestrial radio access base stations. The preliminary experiment confirm the accuracy of AlN film susceptibility analysis and infer a $-6$ dB gain to predicted system sensitivity. Future works include the fabrication of the proposed architecture, and the test in various laboratory and practical scenarios.

\appendices
\section{Reference Sensitivity Power Level}
\subsection{System Sensitivity of Communication Base Stations}\label{Reference_Sensitivity_of_Base_Station}
The reference sensitivity power level $P_{REFSENS}$ is the minimum mean power received at the antenna connector at which a throughput requirement shall be met for a specified reference measurement channel. For evolved universal terrestrial radio access (E-UTRA), the throughput shall be larger than or equal to $95 \%$ of the maximum throughput of the reference measurement channel. The reference sensitivity power levels for Wide Area Base Station (BS), Local Area BS, Home BS and Medium Range BS are shown in Talbe \ref{EUTRA PREFSENS} \cite{etsi2019136}.
\begin{table*}[htbp]
	\centering
	\caption{Reference sensitivity power levels in EUTRA Base Stations.}
	\label{EUTRA PREFSENS}
	\begin{tabular}{|c|c|c|}
		\hline
		& & \\[-6pt]
		Base station types & Bandwidth [MHz] & Reference sensitivity power level [dBm] \\ 
		\hline
		& & \\[-6pt]
		Wide Area Base Station & $5$ & $-101.5$ \\ 
		\hline
		& & \\[-6pt]
		Local Area Base Station & $5$ & $-93.5$ \\
		\hline
		& & \\[-6pt]
		Home Base Station & $5$ & $-93.5$ \\ 
		\hline
		& & \\[-6pt]
		Medium Range a Base Station & $5$ & $-96.5$ \\ 
		\hline
		\multicolumn{3}{|l|}{}\\[-6pt]
		\multicolumn{3}{|l|}{Note: The reference measurement channel is specified in Annex A1-3 in \cite{etsi2019136}.}\\ 
		\hline
	\end{tabular}
\end{table*}

For narrowband internet of thing (NB-IoT) standalone BS or E-UTRA BS with NB-IoT (in-band and/or guard band), NB-IoT throughput shall be larger than or equal to $95\%$ of the maximum throughput of the reference measurement channel. The reference sensitivity power levels for Wide Area BS, Local Area BS, Home BS and Medium Range BS are shown in Talbe \ref{NBIoT PREFSENS} \cite{etsi2019136}.
\begin{table*}[htbp]
	\centering
	\caption{Reference sensitivity power levels in NB-IoT Base Stations.}
	\label{NBIoT PREFSENS}
	\begin{tabular}{|c|c|c|} 
		\hline
		& & \\[-6pt]
		Base station types & Bandwidth [kHz] & Reference sensitivity power level [dBm]\\ 
		\hline
		& & \\[-6pt]
		Wide Area Base Station & $3.75$ & $-133.3$ \\ 
		\hline
		& & \\[-6pt]
		Local Area Base Station & $3.75$ & $-125.3$ \\
		\hline
		& & \\[-6pt]
		Home Base Station & $3.75$ & $-125.3$ \\ 
		\hline
		& & \\[-6pt]
		Medium Range a Base Station & $3.75$ & $-128.3$ \\ 
		\hline
		\multicolumn{3}{|l|}{}\\[-6pt]
		\multicolumn{3}{|l|}{Note: The reference measurement channel is specified in Annex A14-2 in \cite{etsi2019136}.}\\ 
		\hline
	\end{tabular}
\end{table*}

\subsection{System Sensitivity of Atom-Based Sensing System}\label{Reference_Sensitivity_of_Atom_System}
Atom-based measurements have been successfully utilized for magnetometery, time and frequency standards, inertial force sensing6 as well as searches for local Lorentz invariance and intrinsic electric dipole moments of the neutron and electron, amongst others \cite{kumar2017atom}. It is reported that absolute RF electric field sensing is $E_{min} = 5$ $\mu V cm^{-1} Hz^{-1/2}$. The radiation electric field is from a horn antenna, driven by an RF generator. We calculate the equivalent RF generator output signal power as its system sensitivity with power unit Watt.

Assuming that RF generator output signal power is $P_out$, effective radiation area is $A=1 cm^2$, antenna gain is $G_A = 15$ dB and bandwidth is $B=1$ Hz, the electric field strength sensed by the atom system is given by
\begin{equation}
 E_0 = \sqrt{ \frac{2 \langle |\vec{S}|\rangle }{\epsilon_{0} c B}},
\end{equation}
where $\langle |\vec{S}|\rangle = G_A P_{out}/A$ is the energy flow density of electromagnetic wave, $\epsilon_{0}$ is the vacuum dielectric constant and c is the speed of light, respectively.

Accoding to the elctric field sensing sensitivity $E_{min} = 5$ $\mu V cm^{-1} Hz^{-1/2}$, the system sensitivity in the form of power is given by
\begin{equation}
P_{min} = \frac{A}{2 G_A} \epsilon_{0} c E_{min}^2 B \approx 1\times 10^{-15} W = -120 \text{ dBm}.
\end{equation}

\section{Proof of Lemmas and Theorems}
\subsection{Proof of Theorem \ref{theorem_solution_no_damping}}\label{solution_no_damping}
Considering the separation of variables, i.e., $u\left( z,t \right) =Z\left( z \right) \cdot e^{j\omega t}$, and defining the phase velocity $v_a \triangleq \sqrt{\frac{c^D}{\rho}}$ of the vibration module along the z-axis, the wave equation $Z(z)$ is given as follows,
\begin{equation}
	\begin{aligned}
		\omega ^2Z+\frac{c^D}{\rho}\frac{d^2 Z}{dz^2} &=0,\\
		\frac{d^2Z}{dz^2}+\frac{\omega ^2}{c_{v}^{2}}Z &=0,\\
	\end{aligned}
\end{equation}
whose formal solution is
\begin{equation}
	\begin{aligned}
		Z(z) &= A \sin \left( \beta z \right) + B \cos \left( \beta z \right),\\
	\end{aligned}
\end{equation}
where $\beta = {\omega}/{v_a}$ is the amplitude of the wave vector.

According to Maxwell equation $\nabla \cdot D=\rho _{free} = 0$, the electric displacement vector is a variable only depends on time, ${D}(z,t)=D(t)=D_0 e^{j \omega t}$. Substituting boundary condition $T|_{z=0}=T|_{z=L_T}=0$ into the solution, then we can get that $A = {e_{33} D_0}/({\varepsilon ^S c^D \beta})$ and $B = A  [{\cos \left( \beta L_T \right) -1}]/({\sin \left( \beta L_T \right)})$. A complete solution can be given by
\begin{equation}
	u\left( z,t \right) =\frac{e_{33} D_0}{\varepsilon ^S c^D \beta} \frac{\cos \left( \beta \left( L_T-z \right) \right) -\cos \left( \beta z \right)}{\sin \left( \beta L_T \right)} e^{j\omega t}.
\end{equation}
The surface vibration at $z=L_T$ is given by
\begin{equation}\label{surface_solution_no_damping}
	\begin{aligned}
		u\left(L_T, t \right) 
		&=\frac{e_{33} D_0}{\varepsilon ^S c^D \beta} \frac{1 -\cos \left( \beta L_T \right)}{\sin \left( \beta L_T \right)}   e^{j\omega t},\\
		&=\frac{e_{33} D_0}{\varepsilon ^S c^D \beta}\tan \left( \frac{\beta L_T}{2} \right)  e^{j\omega t}.\\
	\end{aligned}
\end{equation}

Integrate the electric field $E$ in the $z$ direction to obtain the expression of the electric potential as follows
\begin{equation}
	\begin{aligned}
		V &=V_0\cdot e^{j\omega t}\\
		&=\int_0^{L_T}{E\left( z,t \right) dz=\int_0^{L_T}{\frac{1}{\varepsilon ^S}\left( D-eS \right) dz}}\\
		&=\frac{D L_T}{\varepsilon ^S}-\frac{e_{33}}{\varepsilon ^S}\left( u\left( L_T,t \right) -u\left( 0,t \right) \right)\\
		&=\left\{ \frac{D_0 L_T}{\varepsilon ^S}-\frac{e_{33}}{\varepsilon ^S}\left[ u\left( L_T,0 \right) -u\left( 0,0 \right) \right] \right\} \cdot e^{j\omega t}\\
		&=\left[ \frac{D_0 L_T}{\varepsilon ^S}-\frac{2e_{33}^2 D_0}{(\varepsilon ^S)^2 c^D\beta}\tan \left( \frac{\beta L_T}{2} \right) \right] \cdot e^{j\omega t},\\
	\end{aligned}
\end{equation}
It is easy to obtain the expression of $D_0$ as follows
\begin{equation}\label{equation_D0}
	D_0 = \frac{V_0 \beta \varepsilon ^S}{ \beta L_T-{2k_{t}^{2}}\tan \left( {\beta L_T / 2} \right)},\\
\end{equation}
where $k_t^2 = e_{33}^2/(c^D \varepsilon^S)$ is electromechanical coupling coefficient, a parameter that characterizes the properties of piezoelectric films. 

Ideally, the series resonance condition corresponds to zero input impedance. The input impedance for thickness excitation case is 
\begin{equation}
Z_{in}=\frac{1}{j\omega C_0}(1-k_t^2\frac{\text{tan}(\beta L_T / 2)}{\beta L_T / 2}),
\end{equation}
where $C_0 = {LW \varepsilon^S}/{L_T}$ is the static capacitance of piezoelectric film \cite{rosenbaum1988bulk}. When $Z_{in}=0$, resonance frequency $\omega_s$ should satisfy
\begin{equation}\label{e1}
\begin{aligned}
k_t^2{\text{tan}(\beta L_T / 2)} &={\beta L_T / 2},\\
\omega_s &= \beta v_a. 
\end{aligned}
\end{equation} 
Near the first pole, the tangent function can be approximated as
\begin{equation}\label{e2}
\tan \left( \frac{\beta L_T}{2} \right) \approx \frac{4\beta L_T}{\pi ^2-\left( \beta L_T \right) ^2}=\frac{\beta L_T}{2k_{t}^{2}}.
\end{equation}
According to Eq. \ref{e1} and Eq. \ref{e2}, the series resonance frequency is given by
\begin{equation}\label{e3}
\omega_s = \frac{v_a}{L_T}\sqrt{\pi ^2-8k_t^2}.
\end{equation}

\subsection{Proof of Theorem \ref{theorem_solution_damped}}\label{solution_damped}
Considering the separation of variables, i.e., $v\left( z,t \right) =V\left( z \right) \cdot e^{j\omega t}$, the wave equation $V(z)$ is given as follows,
\begin{equation}
	c^D\frac{d^2V}{d^2z}+j\omega \eta \frac{d^2V}{dz^2} =-\rho \omega ^2V,
\end{equation}
and the formal solution is 
\begin{equation}
\begin{aligned}
	V\left( z \right) &= A \sin \left( \hat{\beta}z \right) + B\cos \left( \hat{\beta}z \right),\\
\end{aligned}	
\end{equation}
where $\hat{\beta}={\beta}/{\sqrt{1+j\omega \eta/c^D}}\approx \beta\left(1-j/(2Q) \right)$ under the condition for first-order approximation; and the quality factor $Q= c^D/(\eta \omega)$. According to Eq. (\ref{surface_solution_no_damping}), the surface vibration at $z=L_T$ is given by
\begin{equation}\label{equation_u0_damped}
	\begin{aligned}
		u\left( L_T,t \right) &=\frac{e_{33} D_0}{\varepsilon ^S c^D \hat{\beta}}\tan \left( \frac{\hat{\beta}L_T}{2} \right) \cdot e^{j\omega t}.\\
	\end{aligned}
\end{equation}

According to Eq. (\ref{equation_D0}), the amplitude of electric displacement vector is approximated as
\begin{equation}\label{equation_D0_damped}
\begin{aligned}
D_0 &=\frac{V_0\varepsilon ^S}{L_T}\frac{1}{ 1-k_{t}^{2}\frac{\tan \left( \hat{\beta}L_T/2 \right)}{\hat{\beta}L_T/2}}.\\
\end{aligned}
\end{equation}

Based on the series resonance condition as shown in Eq. \ref{e3}, the approximate result is as follows
\begin{equation}\label{e4}
\begin{aligned}
\frac{\tan \left( \frac{\hat{\beta}L_T}{2} \right)}{\hat{\beta}L_T/2}
&\approx \dfrac{\frac{\tan \left( \frac{\beta L_T}{2} \right) -\tan \left( j\frac{1}{2Q}\frac{\beta L_T}{2} \right)}{1+\tan \left( \frac{\beta L_T}{2} \right) \tan \left( j\frac{1}{2Q}\frac{\beta L_T}{2} \right)}}{\frac{\beta L_T}{2}\left( 1-j\frac{1}{2Q} \right)}\\
&\approx \frac{x\left( \frac{1}{k_{t}^{2}}-j\frac{1}{2Q} \right)}{1+j\frac{1}{k_{t}^{2}}\frac{1}{2Q}x^2}\cdot \frac{1}{x\left( 1-j\frac{1}{2Q} \right)}\\
&\approx \frac{1}{k_{t}^{2}}\frac{1}{1+j\frac{1}{k_{t}^{2}}\frac{1}{2Q}x^2}\\
&=\frac{1}{k_{t}^{2}+j\frac{x^2}{2Q}},\\
\end{aligned}
\end{equation}
where $x=\beta L_t / 2$ and $\beta L_T / (4Q) \ll 1$.

Substitute Eq. \ref{e4} into Eq. \ref{equation_D0_damped} and Eq. \ref{equation_u0_damped}, the amplitude of electrical displacement vector is
\begin{equation}
\begin{aligned}
D_0 &\approx \frac{V\varepsilon ^S/L_T}{1-\frac{k_{t}^{2}}{k_{t}^{2}+j\frac{x^2}{2Q}}}
    &=\frac{V\varepsilon ^S}{L_T} \left( 1-j\frac{8Qk_{t}^{2}}{\pi ^2-8k_{t}^{2}} \right),
\end{aligned}
\end{equation}
and the surface vibration equation is 
\begin{equation}
\begin{aligned}
u\left(L_T, t\right) &=\frac{e_{33}D_0}{\varepsilon ^Sc^D\hat{\beta}}\tan \left( \frac{\hat{\beta}L_T}{2} \right) \cdot e^{j\omega t}\\
&=\frac{e_{33}D_0L_T}{2\varepsilon ^Sc^D}\frac{\tan \left( \frac{\hat{\beta}L_T}{2} \right)}{\frac{\hat{\beta}L_T}{2}}\cdot e^{j\omega t}\\
&\approx \frac{e_{33}L_T}{2\varepsilon ^Sc^D}\cdot \frac{V_0\varepsilon ^S}{L_T}\left( 1-j\frac{8Qk_{t}^{2}}{\pi ^2-8k_{t}^{2}} \right) \cdot \frac{1}{k_{t}^{2}+j\frac{\pi ^2-8k_{t}^{2}}{8Q}}\cdot e^{j\omega t}\\
&=\frac{V_0e_{33}}{2c^Dk_{t}^{2}}\cdot \frac{1-j\frac{8Qk_{t}^{2}}{\pi ^2-8k_{t}^{2}}}{1+j\frac{\pi ^2-8k_{t}^{2}}{8Qk_{t}^{2}}}\cdot e^{j\omega t}\\
&\approx -j\frac{4QV_0e_{33}}{c^D\pi ^2}\cdot e^{j\omega t}.\\
\end{aligned}
\end{equation}
 
\subsection{Proof of Theorem \ref{theorem_output_current}}\label{output_current}
Assuming that $r=t={1}/{\sqrt{2}}$ and the phase difference of reflection and transmission can be ignored, the output electric filed can be given by
\begin{equation}
\begin{aligned}
\mathbf{E_{out}} 
	&=\frac{1}{2} \mathbf{E_0} \left( e^{\text{i}\left(\Phi _1 + 2 k_0 x_m\right)}+e^{\text{i}\Phi _2 } \right)\\
	&=\frac{1}{2} \mathbf{E_0} \left( e^{\text{i}\left( -2 k_0 L_N\right)} \left(1 + i 2 k_0 a_s cos(\omega_s t + \varphi_s )\right) \right.\\
	& \quad \left. + e^{\text{i}\left( -2 k_0 L_E\right)} \right).\\
\end{aligned}
\end{equation}
Define the common and differential arm lengths as $\bar{L}= \frac{L_N + L_E}{2}$ and $\Delta L = L_N - L_E$. Noticed the fact that light power $P$ is proportiaonal to the square of electric filed amplitude $|E_0|^2$, the output light power will be as follows,
\begin{equation}
\begin{aligned}
P_{out} &= \frac{\mathbf{E_{out}}\mathbf{E_{out}^*}}{\mathbf{E_0}\mathbf{E_0^*}} P_0 \\
& = P_0 \left[cos(k_0 \Delta L) + k_0 a_s cos(\omega_s t \varphi_s)e^{i(k_0 \Delta L - \pi/2)} \right] \cdot \\
& \quad \quad \left[cos(k_0 \Delta L) + k_0 a_s cos(\omega_s t \varphi_s)e^{i(k_0 \Delta L - \pi/2)} \right]^*\\
& = P_0  \left[ cos^2(k_0 \Delta L) + k_0 a_s sin(2 k_0 \Delta L)cos(\omega_s t + \varphi_s) \right.\\
& \quad \left. + k_0^2 a_s^2 cos^2(\omega_s t + \varphi_s\right].
\end{aligned}
\end{equation}
The output current of photodetector is given by
\begin{equation}
\begin{aligned}
I_{out} &= \alpha P_0  \left[ cos^2(k_0 \Delta L) + k_0 a_s sin(2k_0 \Delta L)cos(\omega_s t + \varphi_s) \right.\\
& \quad \left. + k_0^2 a_s^2 cos^2(\omega_s t + \varphi_s\right].
\end{aligned}
\end{equation}

\subsection{Proof of Theorem \ref{theorem_output_current_with_cavity}}\label{output_current_with_cavity}
Firstly, we use $G_{PRM}$ and $G_{SRM}$ characterise the gain of power cycling cavity and signal cycling cavity in Figure \ref{fig.optical_model_with_cavity}. Then, we can simpify the module to be a Michelson interferometer with two arm cavities, which injection field $\mathbf{E_1} = G_{PRM} \mathbf{E_0}$ and output filed $\mathbf{E_S} = \mathbf{E_{out}}/G_{SRM}$.

Secondly, we analyse the input-output relationship of a two mirror cavity with $ITMN(r_1, t_1)$, $ETMN(r_2, t_2)$ and length $L_N$, the electric field can be given as follows,
\begin{equation}
\begin{aligned}
\mathbf{E_{3}} &= t_1 \mathbf{E_2} + r_1 \mathbf{E_3^{'}}, \\
\mathbf{E_3^{'}} &= r_2 \mathbf{E_3} e^{-i 2 k_0 L_N} e^{i 2 k_0 x_m}, \\
\mathbf{E_2^{'}} &= r_1 \mathbf{E_2} + t_1 \mathbf{E_3^{'}}. \\
\end{aligned}
\end{equation}
The output field $\mathbf{E_2^{'}}$ is given by
\begin{equation}
\mathbf{E_{2}^{'}}=\mathbf{E_{2}}\left(r_1-\frac{r_2 t_1^2 e^{-i 2 k_0 L_N}}{e^{-i 2 k_0 x_m}- r_1 r_2 e^{-i 2 k_0 L_N} }\right),
\end{equation}
where $r_1$, $t_1$, $r_2$ and $t_2$ are the reflection coefficient, transmission coefficient of mirror $ITMN$, and the reflection coefficient, transmission coefficient of mirror $ETMN$, respectively.

Similarly, the output field $\mathbf{E_4^{'}}$ of east arm can be given by
\begin{equation}
\mathbf{E_{4}^{'}}=\mathbf{E_{4}}\left(r_3-\frac{r_4 t_3^2 e^{-i 2 k_0 L_E}}{1 - r_3 r_4 e^{-i 2 k_0 L_E} }\right),
\end{equation}
where $r_3$, $t_3$, $r_4$ and $t_4$ are the reflection coefficient, transmission coefficient of mirror $ITME$, and the reflection coefficient, transmission coefficient of mirror $ETME$, respectively.

Finally, assuming that $r=t={1}/{\sqrt{2}}$ and the phase difference of reflection and transmission can be ignored, the output electric filed can be given by
\begin{equation}
\begin{aligned}
	\mathbf{E_{out}} &= \mathbf{E_S} G_{SRM}\\
	& = (r \mathbf{E_{4}^{'}} + t \mathbf{E_{2}^{'}}) G_{SRM}\\
	& = G_{SRM} \left[\mathbf{E_{4}}\left(r_3-\frac{r_4 t_3^2 e^{-i 2 k_0 L_E}}{1 - r_3 r_4 e^{-i 2 k_0 L_E} }\right)\right.\\
	& \quad  + \left.\mathbf{E_{2}}\left(r_1-\frac{r_2 t_1^2 e^{-i 2 k_0 L_N}}{e^{-i 2 k_0 x_m}- r_1 r_2 e^{-i 2 k_0 L_N} }\right)\right]\\
	& = \frac{1}{\sqrt{2}} G_{PRM} G_{SRM} \mathbf{E_{0}} \left[\left(r_3-\frac{r_4 t_3^2 e^{-i 2 k_0 L_E}}{1 - r_3 r_4 e^{-i 2 k_0 L_E} }\right)\right.\\
	& \quad  + \left.\left(r_1-\frac{r_2 t_1^2 e^{-i 2 k_0 L_N}}{e^{-i 2 k_0 x_m}- r_1 r_2 e^{-i 2 k_0 L_N} }\right)\right].\\
\end{aligned}
\end{equation}

In this work, we choose symmetry parameters for two arm cavities, $r_1=r_3$, $r_2=r_4$, $t_1=t_3$ and $t_2=t_4$. In order to obtain high sensitivity, the reflection coefficient $r_1, r_2$ approximately equal to $1$ and $t_1, t_2 \ll 1$. The output current is given by
\begin{equation}
	\begin{aligned}
		I_{out} & = \frac{1}{2} G_{PRM}^2 G_{SRM}^2 P_0 \left|\left(1-\frac{t_3^2 e^{-i 2 k_0 L_N}}{1 - e^{-i 2 k_0 L_N} }\right)\right.\\
		& \quad  + \left.\left(1-\frac{t_1^2 e^{-i 2 k_0 L_N}}{e^{-i 2 k_0 x_m}- e^{-i 2 k_0 L_N} }\right)\right|^2.\\
	\end{aligned}
\end{equation}
\subsection{Proof of Theorem \ref{theorem_transfer_function}}\label{transfer_function}

Hamiltonian in classical form of POEMS is given by \cite{bagci2014optical}
\begin{equation}
	\begin{aligned}
		H &= \frac{p^2}{2m}+\frac{m\omega _{M}^{2}x^2}{2}+\frac{\phi ^2}{2L}+\frac{q^2}{2C\left( x \right)}+\\
		& \quad 2g\sqrt{\frac{m\omega _M}{C\left( x \right) \omega _{LC}}}xq-qV.
	\end{aligned}
\end{equation}
Under resonance condition, $\omega _M=\omega _{LC}=\omega _r$. The canonical equations are given as follows,
\begin{equation}
	\begin{aligned}
		\frac{\partial H}{\partial p} &=\dot{x},\\
		\frac{\partial H}{\partial x} &=-\dot{p},\\
	\end{aligned} 
\end{equation}
and
\begin{equation}
	\begin{aligned}
		\dot{p} &=-\frac{\partial H}{\partial x}\\
				&=-m\omega _{r}^{2}x-\frac{q^2}{2}\frac{\partial}{\partial x}\left( \frac{1}{C\left( x \right)} \right) -2g\sqrt{m}\frac{\partial}{\partial x}\left( \frac{x}{\sqrt{C\left( x \right)}} \right) q,\\
		\dot{\phi} &=-\frac{\partial H}{\partial q},\\
				&=-\frac{q}{C\left( x \right)}-2g\sqrt{m}\frac{x}{\sqrt{C\left( x \right)}}+V.\\
	\end{aligned}
\end{equation}

Assuming a real system with random disturbance, the Langevin Equations are given by
\begin{equation}
	\begin{aligned}
		\dot{x} &= \frac{p}{m},\\
		\dot{p} &= -m\omega _{r}^{2}x-\frac{q^2}{2}\frac{\partial}{\partial x}\left( \frac{1}{C\left( x \right)} \right) -2g\sqrt{m}\frac{\partial}{\partial x}\left( \frac{x}{\sqrt{C\left( x \right)}} \right) q\\
				&\quad -\Gamma _mp-F,\\
		\dot{q} &=\frac{\phi}{L},\\
		\dot{\phi} &=-\frac{q}{C\left( x \right)}-2g\sqrt{m}\frac{x}{\sqrt{C\left( x \right)}}-\Gamma _{LC}\phi +V,\\
	\end{aligned}
\end{equation}
where $m$ and $L_{LC}$ are equivalent mass of piezoelectric oscillator and inductance in the resonant circuit, respectively; $\omega_{m}$ and $\omega_{LC}$ are resonance frequencies of oscillator and LC cirruit, respectively; $\Gamma_m$ and $\Gamma_{LC}$ are the damping coefficients of the piezoelectric film and the LC resonanct circuit, respectively; and $G$ is the coupling coefficient.

Assume that the equivalent force of the thermal motion $\delta F_{th}$ is the only force. The first order perturbation equations around the equilibrium point $\overline{x}=\overline{p}=\overline{q}=\overline{\phi }=0$, are as follows,
\begin{equation}
	\begin{aligned}
		\delta \dot{x}&=\frac{\delta p}{m},\\
		\delta \dot{p}&=-m\omega _{r}^{2}\delta x-\frac{{\overline{q}}^2}{2}\frac{\partial ^2}{\partial x^2}\left( \frac{1}{C\left( x \right)} \right) _{x=\overline{x}}\delta x\\
		& \quad -\overline{q}\frac{\partial}{\partial x}\left( \frac{1}{C\left( x \right)} \right) _{x=\overline{x}}\delta q - 2\overline{q} g\sqrt{m}\frac{\partial ^2}{\partial x^2}\left( \frac{x}{\sqrt{C\left( x \right)}} \right)_{x=\overline{x}}\delta x\\
		& \quad -2g\sqrt{m}\frac{\partial}{\partial x}\left( \frac{x}{\sqrt{C\left( x \right)}} \right) _{x=\overline{x}}\delta q - \Gamma _m\delta p-\delta F_{th},\\
		\delta \dot{q}&=\frac{\delta \phi}{L},\\
		\delta \dot{\phi}&=-\frac{\delta q}{C\left( \overline{x} \right)}-\overline{q}\frac{\partial}{\partial x}\left( \frac{1}{C\left( x \right)} \right) _{x=\overline{x}}\delta x\\
		& \quad -2g\sqrt{m}\frac{\partial}{\partial x}\left( \frac{x}{\sqrt{C\left( x \right)}} \right) _{x=\overline{x}}\delta x- \Gamma _{LC}\delta \phi +\delta V.\\
	\end{aligned}
\end{equation}
Transforming the analysis from time domain to frequency domain, we can get the equations as follows, 
\begin{equation}
	\begin{aligned}
		-i\Omega \delta x\left( \Omega \right) &=\frac{\delta p\left( \Omega \right)}{m},\\
		-i\Omega \delta p\left( \Omega \right) &=-m\omega _{r}^{2}\delta x\left( \Omega \right) -2g\sqrt{\frac{m}{C\left( \overline{x} \right)}}\delta q\left( \Omega \right) \\
		& \quad -\Gamma _m\delta p\left( \Omega \right) -\delta F_{\text{th}}\left( \Omega \right),\\
		-i\Omega \delta q\left( \Omega \right) &=\frac{\delta \phi \left( \Omega \right)}{L},\\
		-i\Omega \delta \phi \left( \Omega \right) &=-\frac{\delta q\left( \Omega \right)}{C\left( \overline{x} \right)}-2g\sqrt{\frac{m}{C\left( \overline{x} \right)}}\delta x\left( \Omega \right) \\
		& \quad -\Gamma _{LC}\delta \phi \left( \Omega \right) + \delta V\left( \Omega \right).\\
	\end{aligned}
\end{equation}
Redefine the coupling coefficient $G=2g\sqrt{\frac{m}{C\left( \overline{x} \right)}}=2g\sqrt{\frac{mL_T}{\epsilon ^S WL}}$, the equations in frequency domain are given as follows,
\begin{equation}
	\begin{aligned}
		-i\Omega \delta x\left( \Omega \right) &=\frac{\delta p\left( \Omega \right)}{m},\\
		-i\Omega \delta p\left( \Omega \right) &=-m\omega _{r}^{2}\delta x\left( \Omega \right) -\Gamma _{\text{m}}\delta p\left( \Omega \right) -G\delta q\left( \Omega \right) -\delta F_{\text{th}}\left( \Omega \right),\\
		-i\Omega \delta q\left( \Omega \right) &=\frac{\delta \phi \left( \Omega \right)}{L},\\
		-i\Omega \delta \phi \left( \Omega \right) &=-\frac{\delta q\left( \Omega \right)}{C\left( \overline{x} \right)}-\Gamma _{\text{LC}}\delta \phi \left( \Omega \right) -G\delta x\left( \Omega \right) +\delta V\left( \Omega \right).\\
	\end{aligned}
\end{equation}

According to the above equation, the response of the system to force or voltage signal excitation can be calculated. For piezoelectric oscillator and LC circuit, two parameters are defined as follows,
\begin{equation}
	\begin{aligned}
		\chi _{\text{m}}\left( \Omega \right) &=\frac{1}{m\left( \Omega _{\text{m}}^{2}-\Omega ^2-i\Omega \Gamma _{\text{m}} \right)},\\
		\chi _{\text{LC}}\left( \Omega \right) &=\frac{1}{L\left( \Omega _{\text{LC}}^{2}-\Omega ^2-i\Omega \Gamma _{\text{LC}} \right)}.\\
	\end{aligned}
\end{equation}
Again, the equations in frequency domain can be written as follows,
\begin{equation}
	\begin{aligned}
		&\chi _m\left( \Omega \right) \left( -\delta F_{th}\left( \Omega \right) +G\chi _{LC}\left( \Omega \right) \delta V\left( \Omega \right) \right) \\
		&\quad =\left( 1-G^2\chi _m\left( \Omega \right) \chi _{LC}\left( \Omega \right) \right) \delta x\left( \Omega \right),\\
		&\delta x\left( \Omega \right) =\left( \chi _m\left( \Omega \right) ^{-1}-G^2\chi _{LC}\left( \Omega \right) \right) ^{-1}\\ &\quad \left( -\delta F_{th}\left( \Omega \right) +G\chi _{LC}\left( \Omega \right) \delta V \right).\\
	\end{aligned}
\end{equation}

Defining $\chi _{m}^{eff}\left( \Omega \right) \triangleq \left( \chi _m\left( \Omega \right) ^{-1}-G^2\chi _{LC}\left( \Omega \right) \right) ^{-1}$, the transfer function of the system in frequency domain is given by
\begin{equation}
	\begin{aligned}
	\delta x\left( \Omega \right) &=\chi _{m}^{eff}\left( -\delta F_{th}\left( \Omega \right) +G\chi _{LC}\left( \Omega \right) \delta V\left( \Omega \right) \right),\\
	\delta \varphi \left( \Omega \right) &=2k\chi _{m}^{eff}\left( -\delta F_{th}\left( \Omega \right) +G\chi _{LC}\left( \Omega \right) \delta V\left( \Omega \right) \right) \\
	& \quad +\delta \varphi _{im}\left( \Omega \right),\\
	\end{aligned}
\end{equation}
where $\delta \varphi =2k\delta x\ \left( k=2\pi /\lambda \right) $ represents the phase variation to $\delta x$.

Moreover, considering the light force of the optical cavity in a real system, the expression of $\chi_{m}\left(\omega\right)$ is modified as follows,
\begin{equation}\label{equation_chi_eff}
	\begin{aligned}
		\chi _{m,\text{eff}}^{-1}\left( \omega \right) &=\chi _{m}^{-1}\left( \omega \right) +\Sigma \left( \omega \right), \\
		\Sigma \left( \omega \right) &=2m_{\text{eff}}\Omega _mg^2\left\{ \frac{1}{\left( \Delta +\omega \right) +i\kappa /2}+\frac{1}{\left( \Delta -\omega \right) -i\kappa /2} \right\},\\
	\end{aligned}
\end{equation}
where $\Delta$ and $\kappa$ represent the laser detuning to the optical cavity and attenuation coefficient, respectively \cite{aspelmeyer2014cavity}.

Then, we introduce two parameters $\delta \Omega _m$ and $\Gamma _{opt}$ as follows,
\begin{equation}
	\begin{aligned}
		\delta \Omega _m\left( \omega \right) &=g^2\frac{\Omega _m}{\omega}\left[ \frac{\Delta +\omega}{\left( \Delta +\omega \right) ^2+\kappa ^2/4}+\frac{\Delta -\omega}{\left( \Delta -\omega \right) ^2+\kappa ^2/4} \right],\\
		\Gamma _{\text{opt}}\left( \omega \right) &=g^2\frac{\Omega _m}{\omega}\left[ \frac{\kappa}{\left( \Delta +\omega \right) ^2+\kappa ^2/4}-\frac{\kappa}{\left( \Delta -\omega \right) ^2+\kappa ^2/4} \right],\\
	\end{aligned}
\end{equation}
where $\delta \Omega _m$ and $\Gamma _{opt}$ represent frequency shift and damping coefficient variation caused by light incident, respectively. Substituting them into Eq. (\ref{equation_chi_eff}),
\begin{equation}
	\begin{aligned}
		\Sigma \left( \omega \right) &\equiv m_{\text{eff}}\omega \left[ 2\delta \Omega _m\left( \omega \right) -i\Gamma _{\text{opt}}\left( \omega \right) \right], \\
		\chi _{m,\text{eff}}^{-1}\left( \omega \right) & = m_{eff}\left[ \Omega _{m}^{2}+2\omega \delta \Omega _m\left( \omega \right) -\omega ^2\right]\\
		& \quad -i\omega m_{eff} \left[ \Gamma _m+\Gamma _{\text{opt}}\left( \omega \right) \right].\\
	\end{aligned}
\end{equation}

Under different cavity attenuation sizes $\kappa$, the form of frequency detuning is not exactly the same. The frequency drift introduced by the optical field is called the optical spring effect, which disappears under certain detuning conditions. Moreover, the radiation pressure in the cavity will produce equivalent cooling and amplification effects on the motion of the end mirror (in our system, it is represented by piezoelectric film).

\subsection{Proof of Theorem \ref{theorem_film_thermal_noise}}\label{film_thermal_noise}
In our system, the response of piezoelectric film to RF signals can be approximated as a damped harmonic oscillator model. The vibration equation is given by
\begin{equation}
	m_{\text{eff}}\frac{dx^2\left( t \right)}{dt^2}+m_{\text{eff}}\Gamma _m\frac{dx\left( t \right)}{dt}+m_{\text{eff}}\Omega _{m}^{2}x\left( t \right) =\eta _{ex}\left( t \right).
\end{equation}
In the equilibrium, the environment exerts a disturbing force $\eta_{ex}$ on the damped harmonic oscillator. Defining $\eta \left( t \right) \triangleq \eta _{ex}\left( t \right) /m_{eff}$, the vibration equation is simpified as follows,
\begin{equation}
	\frac{dx^2\left( t \right)}{dt^2}+\Gamma _m\frac{dx\left( t \right)}{dt}+\Omega _{m}^{2}x\left( t \right) =\eta \left( t \right) .
\end{equation}

According to the white noise hypothesis, $\eta \left( t \right) $ satisfies the following equations,
\begin{equation}
\left< \eta \left( t \right) \right> =0,\ \left< \eta \left( t \right) \eta \left( \tau \right) \right> =2\alpha \delta \left( t-\tau \right). 
\end{equation}
A solution is given by
\begin{equation}
	\begin{aligned}
		x\left( t \right) &=a_{10}\text{e}^{\mu _1t}+a_{20}\text{e}^{\mu _2t}\\
		& \quad + \frac{1}{\mu_1-\mu_2}\int_0^t{\left[ \text{e}^{\mu _1\left( t-t^\prime \right)}-\text{e}^{\mu _2\left( t-t^\prime \right)} \right]}\eta \left( t^\prime \right) \text{d}t^\prime, \\
	\end{aligned}
\end{equation}
where $\mu _1$ and $\mu _2$ are the two solutions to $\mu ^2+\Gamma _m\mu +\Omega _{m}^{2}=0$, respectively. Under weak stationary condition, the correlation function of simple harmonic noise is given by \cite{bao2009random}
\begin{equation}
\left< x\left( t \right) x\left( \tau \right) \right> =\frac{\alpha}{\mu _{1}^{2}-\mu _{2}^{2}}\left[ \frac{1}{\mu _1}\text{e}^{\mu _1|t-\tau |}-\frac{1}{\mu _2}\text{e}^{\mu _2|t-\tau |} \right] .
\end{equation}

Fourier transform is applied to the above equation to obtain the simple harmonic noise power spectrum as follows,
\begin{equation}
	\begin{aligned}
		S\left( \omega \right) &=\ 2\alpha _{ex}|\chi _m\left( \omega \right) |^2,\\
	\end{aligned}
\end{equation}
where $\alpha _{ex}=\alpha m_{eff}^{2}$ represents the noise strength.



\ifCLASSOPTIONcaptionsoff
  \newpage
\fi

\bibliographystyle{./IEEEtran}
\bibliography{./mybib}

\end{document}